\documentclass[12pt,dvips]{article}
\usepackage{amsfonts}
\usepackage{amsmath}
\usepackage{amssymb}
\usepackage{appendix}
\usepackage[small]{caption}
\usepackage{cite}
\usepackage{color}
\usepackage{enumerate}
\usepackage{graphicx}
\usepackage{hhline}
\usepackage{longtable}
\usepackage{multirow}
 \usepackage{rotating}
\usepackage{subfigure}

\def\sla#1{\setbox0=\hbox{$#1$}\dimen0=\wd0
  \setbox1=\hbox{/} \dimen1=\wd1 \ifdim\dimen0>\dimen1
  \rlap{\hbox to \dimen0{\hfil/\hfil}} #1
  \else
  \rlap{\hbox to \dimen1{\hfil$#1$\hfil}}
  /   \fi}

\def\bea{\begin{eqnarray}}
\def\eea{\end{eqnarray}}
\definecolor{brown}{rgb}{0.5,0.2,0.0}

\begin{document}

\def\thefootnote{\fnsymbol{footnote}}

\begin{flushright}
  {\tt APCTP Pre2012-013}\\
  {\tt KIAS-P12027}\\
  {\tt IFT-UAM/CSIC-12-84}\\
  % Last modified by CSS on Aug 1st
  %Last modified by CSS on \today\\
  % Last modified by EJC on Aug. 8\\
\end{flushright}

\begin{center}
  {\bf {\Large Peccei-Quinn NMSSM in the light of 125 GeV Higgs
    }}
\end{center}

\medskip

\begin{center}{\large
    Kyu~Jung~Bae$^{a,}$\footnote{Email: kyujung.bae@kaist.ac.kr},
    Kiwoon~Choi$^{a,b,}$\footnote{Email: kchoi@kaist.ac.kr},
    Eung~Jin~Chun$^{b,}$\footnote{Email: ejchun@kias.re.kr},
    Sang~Hui~Im$^{a,}$\footnote{Email: shim@muon.kaist.ac.kr},\\[0.2cm]
    Chan~Beom~Park$^{c,}$\footnote{Email: chanbeom.park@csic.es}, and
    Chang~Sub~Shin$^{d,}$\footnote{Email: csshin@apctp.org}
  }
\end{center}

\begin{center}
  {\em $^a$Department of Physics, KAIST, Daejeon 305-701,
    Korea}\\[0.2cm]
  {\em $^b$Korea Institute for Advanced Study, Seoul 130-722,
    Korea}\\[0.2cm]
  {\em $^c$Instituto de F\'isica Te\'orica UAM/CSIC,\\
    Universidad Aut\'onoma de Madrid, Cantoblanco, 28049 Madrid,
    Spain}\\[0.2cm]
  {\em $^d$Asia Pacific Center for Theoretical Physics, \\
    Pohang, Gyeongbuk 790-784, Korea}\\[0.2cm]
\end{center}

\bigskip\bigskip

\centerline{\bf ABSTRACT}
\noindent
We study the phenomenology of the Peccei-Quinn invariant extension
of the next-to-minimal supersymmetric standard model (NMSSM) in view
of the recent discovery of a 125 GeV Higgs boson. The minimal model
having no quadratic and cubic terms of the NMSSM singlet field
predicts a light singlino-like lightest supersymmetric particle (LSP).
The model is strongly constrained by the Higgs invisible decay
and the dark matter characteristic of the LSP,
while some constraints can be relaxed by assuming that the saxion, the CP-even
companion of the axion in the Peccei-Quinn sector, causes a
late-time entropy production diluting the thermal LSP density.
The collider signal of the model contains multi-jet and $h/W/Z$
plus missing energy, which can be discovered in the early
stage of the 14 TeV LHC running.

\medskip
\noindent
% {\small {\sc Keywords}:
% Hadronic Colliders, Supersymmetry Phenomenology}

\newpage

% -------------------------------------------------------------
\renewcommand{\thefootnote}{\arabic{footnote}}
\setcounter{footnote}{0}

\section{Introduction}

The strong CP problem can be nicely resolved in a supersymmetric
(SUSY) standard model, which allows an extended Higgs sector to
implement the Peccei-Quinn (PQ) symmetry~\cite{Kim08}. After
integrating out the heavy PQ sector around $v_{\rm PQ} =
10^9 - 10^{12}$ GeV, and freezing the axion supermultiplet at their
vacuum values, the low energy theory can be reduced to the
conventional minimal supersymmetric standard model
(MSSM)~\cite{Kim83}, or to the next-to-minimal supersymmetric
standard model (NMSSM)~\cite{Ellwanger:2009dp} as proposed
recently in \cite{Jeong11,Kim12}.

In the PQ-symmetric NMSSM (PQ-NMSSM), the low-energy theory is
generically described by the superpotential,
\begin{equation}
  W = \lambda S H_u H_d + \mu_S^2 S + {1\over2} \mu^\prime_S S^2,
\end{equation}
and corresponding soft-breaking terms in the scalar potential,
where $\mu_S$ and $\mu^\prime_S$ at the TeV scale can arise as a
function of $v_{\rm PQ}$ after the PQ symmetry breaking
at the scale $v_{\rm PQ}$.

In this paper, we investigate the phenomenology of the PQ-NMSSM in
light of the recent LHC results on the Higgs boson search~\cite{Higgs11, Higgs12}.
In its minimal form~\cite{Jeong11}, the model contains only the $\mu_S^2$ term predicting
a very light singlino. Because of this, the minimal model is severely constrained by
the consideration of the Higgs invisible decay and the dark matter
property of the singlino-like lightest supersymmetric particle (LSP)
in the standard cosmology. A sizable NMSSM contribution to
the 125 GeV Higgs boson mass can be obtained with $\tan\beta\approx
1$, which makes the LSP heavy enough to forbid the invisible Higgs
decay. However, the recent XENON100 result on the direct detection of
dark matter~\cite{XENON:2012nq} excludes almost all the LSP mass
region, although the LSP annihilation by the Higgs resonance effect
can reduce the dark matter relic density significantly. For larger
$\tan\beta$, the LSP gets lighter to open the Higgs invisible decay
channel. In this case, the coupling between the Higgs boson and the
LSP can be made small by a cancellation.
Considering the charged Higgs boson mass bound from the LHC, the Higgs
invisible decay branching fraction can be smaller than 0.1 for
$\tan\beta\gtrsim9$. For such a large $\tan\beta$, the NMSSM
contribution to the Higgs boson mass becomes negligible. As the light
singlino being the LSP couples very weakly to the quarks and leptons,
its thermal relic density overcloses the universe.
We will argue that the difficulties of the thermal LSP dark matter can
be circumvented by a late-time entropy production by the saxion field
which is inherent in the PQ-NMSSM, and still a correct amount of dark matter
can be provided by either axion or non-thermal LSP. Analyzing the generic collider
signatures of multi-jet and $h/W/Z$ plus missing energy for two
chosen benchmark points with small and large $\tan\beta$, we find that
a 5$\sigma$ discovery can be achieved in the early stage of the 14 TeV
LHC running.

This paper is organized as follows. In Sec.~\ref{sec:PQNMSSM}, we
provide a general description of PQ-NMSSM. Then, we analyze the Higgs
and the neutralino sectors of the minimal PQ-NMSSM.  In
Sec.~\ref{sec:Higgs_pheno}, we study the phenomenology of a 125 GeV
Higgs boson and a light singlino to constrain the model parameter
space from invisible decays of the $Z$ and the Higgs boson, and the
direct production of light neutralinos at LEP II.
Analyzing the dark matter property of the LSP as a mixture of the
singlino and Higgsino, the minimal model will be tightly constrained
in Sec.~\ref{sec:DM}. In Sec.~\ref{sec:collider}, we analyze the
collider signatures of the PQ-NMSSM at the LHC. We conclude in
Sec.~\ref{sec:concl}.

\section{Peccei-Quinn Symmetric NMSSM}
\label{sec:PQNMSSM}

\subsection{Model of PQ symmetry breaking}
\label{sec:PQ}

The U(1)$_{\rm PQ}$ symmetry, which solves the strong CP problem of
the standard model (SM), should be spontaneously broken between $10^9$
and $10^{12}$ GeV. In SUSY models, it can be realized
by introducing PQ charged but SM singlet chiral superfields \bea
X_I= \phi_I
+ \sqrt{2}\theta\widetilde a_I + \theta^2 F^{X_I}, \eea
whose scalar components are stabilized at an intermediate scale,
$\langle|\phi_I|\rangle \sim v_{\rm PQ}\sim 10^9-10^{12}$ GeV.
Then, the QCD axion corresponds to a linear combination of the axial
components of $\phi_I$. Such a large vacuum value,
compared to the weak scale, can be easily obtained
if $\langle |\phi_I|\rangle$ are determined by the interplay of
soft SUSY breaking terms and the $F$-term scalar potential suppressed by
a cut-off scale.
In order to write down the higher dimensional term, we introduce two
PQ-charged chiral superfields and the superpotential,
\bea\label{PQmodel}
W_{\rm PQ} = \frac{\kappa_{_{\rm PQ}}}{M_{\rm Pl}}
X_1^n X_2^{4-n} \quad (n=1,\,2),\eea
where $M_{\rm Pl}\simeq 2.4\times 10^{18}$ GeV is the reduced Planck
mass. The PQ symmetry is realized as
\bea
{\rm U(1)}_{\rm PQ}:\ X_1\rightarrow e^{-iq_{X_1}\alpha} X_1,\quad
X_2\rightarrow e^{-iq_{X_2} \alpha} X_2,\quad
(nq_{X_1}+(4-n)q_{X_2}=0).\eea
When one includes the soft SUSY breaking terms,
the scalar potential is
\bea
V_{\rm PQ} &=& m_{X_1}^2 |\phi_1|^2 + m_{X_2}^2 |\phi_2|^2 +
\left(\frac{A_{_{\rm PQ}}\kappa_{_{\rm PQ}}}{M_{\rm Pl}} \phi_1^n
  \phi_2^{4-n} + {\rm h.c.}\right) \nonumber\\ &&  +
\frac{\kappa_{_{\rm PQ}}^2}{M_{\rm Pl}^2} \left( n^2
|\phi_1|^{2(n-1)} |\phi_2|^{2(4-n)} + (4-n)^2 |\phi_1|^{2n}
|\phi_2|^{2(3-n)}\right).
\eea
All the soft parameters $|m_{X_1}|\sim |m_{X_2}|
\sim |A_{_{\rm PQ}}|$ are of the order of $m_{\rm soft} ={\cal
  O}(10^2 - 10^3)$ GeV. In the case of $n=2$, the additional $\mathbb{Z}_2$
symmetry ($X_I\rightarrow(-1)^{I}X_I$) is needed to prevent $M_X X_1X_2$
term in the superpotential.
% The detailed vacuum structure is not a scope
% of this paper, however,
The PQ symmetry breaking scale can be estimated as \bea \label{PQVEV1}
v_{\rm PQ}=\sqrt{\frac{1}{2}\Big(q_X^2\langle
  |\phi_1|^2\rangle + q_Y^2\langle|\phi_2|^2\rangle\Big)}
\sim \langle|\phi_1|\rangle \sim \langle|\phi_2|\rangle
\sim \sqrt{m_{\rm soft} M_{\rm Pl}/\kappa_{_{\rm PQ}}},
\eea and it is naturally lying on the axion window for a
moderate value of $\kappa_{_{\rm PQ}}$.
Non-zero auxiliary $F$-components of $X_1$ and $X_2$ are also
developed. In canonical basis of the superfields,
they are given as \bea \label{PQVEV2}
\left\langle \frac{F^{X_1}}{\phi_1}\right\rangle&=& -
\left\langle \frac{1}{\phi_1}\frac{\partial W_{\rm PQ}^\ast}
  {\partial \phi_1^\ast}\right\rangle=-\left\langle \frac{n
    W_{\rm PQ}^\ast}{|\phi_1|^2}\right\rangle\sim
\frac{\kappa_{_{\rm PQ}}v_{\rm PQ}^2}{M_{\rm Pl}}
\sim m_{\rm soft}, \nonumber\\
\left\langle\frac{F^{X_2}}{\phi_2}\right\rangle&=&
-\left\langle\frac{1}{\phi_2}\frac{\partial
    W_{\rm PQ}^\ast}{\partial \phi_2^\ast}\right\rangle=
-\left\langle\frac{(4-n)W^\ast_{\rm PQ}}{|\phi_2|^2}\right\rangle \sim
\left\langle \frac{F^{X_1}}{\phi_1}\right\rangle\sim m_{\rm soft}.
\eea
After fixing the vacuum values, the scalar fields can be decomposed as
$\phi_I = \langle|\phi_I|\rangle +\left(s_I + i a_I\right)/\sqrt{2}$.
Masses of the PQ fields ($s_I$, $a_I$, and $\widetilde a_I$)
are generically of order of $m_{\rm soft}$ except the mass of the
QCD axion $a_{\rm QCD} = \sum_I q_I\langle|\phi_I|\rangle
a_I/\sqrt{2}v_{\rm PQ}$ of order $\sqrt{m_q\Lambda_{\rm QCD}^3}/v_{\rm PQ}$,
where $m_q$ is a light quark mass.
The PQ charged particles with masses of ${\cal O}(m_{\rm soft})$
can play a important role in cosmology,
which will be addressed in Sec.~\ref{sec:DM}.

The axion solution to the strong CP problem can be realized by
extending the Higgs sector through a superpotential term, $ X_1^2
H_u H_d /M_{\rm Pl}$ with $q_{H_uH_d} = - 2 q_{X_1}$ \cite{Kim83}. A
simple consequence of this extension is that a bare Higgs-bilinear
term is forbidden by the PQ symmetry and the right size of the
$\mu$-parameter is generated, \bea \mu_0=
\frac{\langle\phi_1^2\rangle}{M_{\rm Pl}} \sim \frac{v_{\rm
PQ}^2}{M_{\rm Pl}}\sim m_{\rm soft}. \eea
 The low energy theory after the PQ symmetry breaking could also
 be of the NMSSM type \cite{Jeong11,Kim12} where the $\mu_0$ term is extended to a
 scalar field $S$ and the Higgs boson mass at 125 -- 126 GeV could be realized
 with less fine-tuning \cite{Ellwanger:2011na, Ross:2012gn}. In the following subsection,
we will describe how the PQ symmetry is incorporated into the
NMSSM setup.

\subsection{General PQ-NMSSM}

In the NMSSM, the SM singlet superfield $S$ couples to the Higgs
fields with a renormalizable term $\lambda S H_u H_d$ in the
superpotential. Then, $S$ should have the PQ charge, $q_S= -q_{H_u
H_d}$.
Below the PQ symmetry breaking scale, %(around TeV scale),
the effective theory will be described by the general NMSSM \cite{Ross:2012gn}:
\bea
W_{\rm eff}= (\mu_0+ \lambda S) H_u H_d +
\mu_S^2 S + \frac{1}{2} \mu^\prime_S S^2 + \frac{1}{3}
\kappa S^3,
\eea
where $\mu_0$, $\mu_S^2$, $\mu^\prime_S$, and $\kappa$ are the
effective parameters determined by interactions between $S$ and $X_I$.
Obviously, they should be vanishing in the PQ symmetric limit,
$v_{\rm PQ}\rightarrow 0$.
In order for $S$ to be a light degree of freedom that survives around
TeV scale, $\mu_0,\,\mu_S,\,\mu_S^\prime \lesssim {\cal O}(m_{\rm
  soft})$, and corresponding soft SUSY breaking parameters are
around $m_{\rm soft}$ as well. Note that $\mu_0$ and the corresponding
soft SUSY-breaking parameter ($\theta^2$ component of the spurion field $\mu_0$) can be always rotated away
by the holomorphic field redefinition $ S \rightarrow S - \mu_0/\lambda $. So we can set $ \mu_0=0 $ without loss of
generality. Now
it is expected that sizable values of $\mu_S^2$ and $\mu_S^\prime$
can be obtained by the PQ symmetry breaking model.
We will study explicitly how such PQ symmetry breaking parameters can
be generated for the models given in Sec.~\ref{sec:PQ}.% \footnote{
% In gravity mediation, the general NMSSM with
% $\mu_0\sim \mu_S\sim \mu_S'={\cal O}(m_{3/2})$, and $\kappa={\cal O}(1)$
% can be obtained assuming that $S$ and $H_u H_d$ have nontrivial charges
% under the pseudo-anomalous discrete R-symmetry\cite{Lee:2011dya}.
% Since $U(1)_{\rm PQ}$ is not a R-symmetry, the breaking scale of
% $U(1)_{\rm PQ}$ is not directly related with the gravitino mass.
% In our discussion, we do not choose a particular SUSY breaking mediation scheme.}

For the PQ symmetry breaking model~(\ref{PQmodel})
with $n=1$ ($W_{\rm PQ}=\kappa_{_{\rm PQ}}
X Y^3/M_{\rm Pl}$), the relevant PQ invariant
superpotential  is
\bea
\label{PQ-NMSSM}
W=\left(\frac{X_1^2}{M_{\rm Pl}}+ \lambda S\right) H_u H_d    +\cdots,
\eea
where dots denote highly suppressed terms that do not contribute to
the Higgs phenomenology.
The non-trivial $\mu_S^2$ can be obtained from a PQ invariant
higher dimensional term in the K\"ahler potential, \bea\label{PQKahler}
K= \frac{\kappa_{XS}}{M_{\rm Pl}}X_1^{\ast 2} S + {\rm h.c.}.
\eea
At low energy, $\kappa_{XS}$ and $X_I$
can be regarded as SUSY breaking spurion superfields
with $\kappa_{XS}=\kappa_{0}+ \theta^2 \kappa_Fm_{\rm soft}
+\bar\theta^2 \kappa_{\bar F}m_{\rm soft}+\theta^2\bar\theta^2
\kappa_D m_{\rm soft}^2$,
and $X_I=\langle\phi_I\rangle(1+\theta^2\langle F^{X_I}/\phi_I\rangle)$.
$\kappa_{0}$, $\kappa_F$, $\kappa_{\bar F}$, $\kappa_D$
are all ${\cal O}(1)$ constants with a reasonable assumption that
they and the soft SUSY breaking terms for the superpartners
of the SM fields have the same origin.
Then, by (\ref{PQVEV1}) and (\ref{PQVEV2}),
\bea
\mu_S^2 = \left(\kappa_{\bar F} m_{\rm soft}
  +2\left\langle \frac{ F^{X_1^\ast}}{\phi_1^\ast} \right\rangle
\right)\left(\frac{\langle\phi_1^{\ast 2}\rangle}{M_{\rm Pl}}\right)
\sim \left(\frac{v_{\rm PQ}^2}{M_{\rm Pl}}\right)^2\sim m_{\rm soft}^2,
\eea but still $\mu_S^\prime$ and $\kappa$ are suppressed.
Again note that the superpotential term $X_1^2 H_u H_d/M_{\rm Pl}$ in
(\ref{PQ-NMSSM}) can be removed by the holomorphic field redefinition
$S \rightarrow S-X_1^2/\lambda M_{\rm Pl}$ without loss of generality.
Then, the low energy effective superpotential corresponds to
the minimal type of PQ-NMSSM~\cite{Panagiotakopoulos:2000wp},
\bea\label{nMSSM}
W_{\rm eff} = \lambda S H_u H_d + \mu_S^2 S.
\eea

As for the model with $n=2$ ($W_{\rm PQ}=\kappa_{_{\rm PQ}}
X^2 Y^2/M_{\rm Pl}$), the same superpotential (\ref{PQ-NMSSM}) and
K\"ahler potential (\ref{PQKahler}) are allowed so that the
sizable $\mu_0$, $\mu_S^2$ are generated. As discussed in Sec.~\ref{sec:PQ},
there is an additional $\mathbb{Z}_2$ symmetry $X_I\rightarrow (-1)^I X_I$,
$S\rightarrow -S$, $H_uH_d\rightarrow - H_uH_d$ at renormalizable level,
in order to insure that  terms like $X_1 X_2$, $ X_2^2S$ are absent.
Since the $\mathbb{Z}_2$ symmetry is explicitly broken by the term $X_1^2 H_u
H_d/M_{\rm Pl}$, a cosmologically dangerous domain wall is not produced.
Besides, the $\mathbb{Z}_2$ breaking tadpole induced by
supergravity loop corrections are suppressed due to the PQ symmetry.
This model is more complicated than the model with $n=1$. However,
a sizable $\mu_S^\prime$ term can be generated if there are matter
superfields $Z_1$, $Z_2$ whose masses are given by
$\langle |\phi_1|\rangle $ from a superpotential term $ X_1 Z_1 Z_2 $
with the PQ charges $q_{Z_1}=3q_{X_1}$, $q_{Z_2}=-4q_{X_1}$.
Under the $\mathbb{Z}_2$ symmetry, $Z_I\rightarrow (-1)^IZ_I$.
Then, the following superpotentials are allowed
% \bea
{\allowdisplaybreaks \begin{align}
\Delta W ={} & \lambda_{XZ} X_1 Z_1 Z_2 +
\lambda_{ZS} Z_2 S^2 + \frac{1}{M_{\rm Pl}}
\Big( \kappa_{XZ} X_2^3 Z_1 + \kappa_{XZS} X_1^2 Z_2 S \Big)+
\frac{\lambda_{XS}}{M_{\rm Pl}^2}X_2^4S+\cdots.
\label{Model2}
\end{align}
% \eea
Since the masses of $Z_1$ and $Z_2$ are of the order of $v_{\rm PQ}$,
they should be integrated out, at the PQ symmetry breaking scale, by the superfield
equations of motion:}
\bea\label{Integrating}
\frac{\partial W}{\partial Z_1}\simeq \frac{\partial W}{\partial Z_2}
\simeq 0.
\eea By substituting the solutions of (\ref{Integrating}) to  $Z_I$ of (\ref{Model2}),
$\Delta \mu_S^2$ and $\mu^\prime_S$ are generated as
\bea
\Delta\mu_S^2 &=&\left(\frac{\kappa_{XZS}\kappa_{XZ}\langle\phi_1\rangle}
  {\lambda_{XZ}\langle\phi_2\rangle}+ \lambda_{XS}\right)
\left(\frac{\langle\phi_2^2\rangle}{M_{\rm Pl}}\right)^2
\sim\left(\frac{v_{\rm PQ}^2}{M_{\rm Pl}}\right)^2\sim m_{\rm soft}^2,
\nonumber\\
\mu^\prime_S &=&\left(\frac{\lambda_{ZS}\kappa_{XZ}\langle\phi_2\rangle}
  {\lambda_{XZ}\langle\phi_1\rangle}\right)
\left(\frac{\langle \phi_2^2\rangle}
  {M_{\rm Pl}}\right)\sim\frac{v_{\rm PQ}^2}{M_{\rm Pl}} \sim m_{\rm soft}.
\eea
This case corresponds to % the so-called S-MSSM model,
the singlet extension of the MSSM~\cite{Delgado:2010uj},
\bea\label{S-NMSSM}
W_{\rm eff} = \lambda S H_u H_d + \mu_S^2 S + \mu_S^\prime S^2,
\eea
while $\kappa$ still suppressed and $\mu_0$ rotated away. The suppressed $\kappa$ is a generic consequence of the PQ extension of the NMSSM.

The PQ sector contributions to the soft SUSY-breaking parameters
for Higgs and singlet sector are coming from $\theta^2$ ($\bar\theta^2$)
component of the spurion superfields $X_I=\langle\phi_I\rangle
(1+\theta^2\langle F^{X_I}/\phi_I\rangle)$ ($X^\ast_I$),
which are the same order of the MSSM soft SUSY-breaking
parameters, $m_{\rm soft}$.

\subsection{Minimal PQ-NMSSM}

In this subsection, we investigate the phenomenological
consequences of the minimal PQ-NMSSM \cite{Jeong11} whose
low-energy effective superpotential takes the form of
(\ref{nMSSM}) in addition to the usual MSSM Yukawa superpotential terms.
The new superfield $S$ is singlet under the SM gauge group,
%$SU(3)_c \times SU(2)_L \times U(1)_Y$,
and it can acquire the vacuum expectation value (VEV) to give the natural
size of the $\mu$ term of the electroweak (EW) scale.

\subsubsection{Higgs sector}
\label{sec:Higgs}

Here, we describe the Higgs scalar potential, its
vacuum structure, and the mass spectra around the vacuum. The
Higgs scalar potential consists of the following $F$- and $D$-term
contributions from the superpotential (\ref{nMSSM}), and soft
SUSY-breaking terms,
{\allowdisplaybreaks\begin{align}
  V_F = {} & \left|\lambda(H_u^+ H_d^- - H_u^0 H_d^0) + \mu_S^2\right|^2 + \lambda^2 |S|^2 \left(|H_u^0|^2+|H_u^+|^2+|H_d^0|^2+|H_d^-|^2\right), \\
  V_D = {} & \frac{g_1^2 + g_2^2}{8}\left(|H_u^0|^2+|H_u^+|^2-|H_d^0|^2-|H_d^-|^2\right)^2 +\frac{g_2^2}{2}\left|H_u^+{H_d^0}^\ast + H_u^0 {H_d^-}^\ast \right|^2, \\
  V_S = {} & m_{H_u}^2\left(|H_u^0|^2+|H_u^+|^2\right)+m_{H_d}^2\left(|H_d^0|^2+|H_d^-|^2\right)+m_S^2|S|^2 \nonumber \\
  &+\left[\lambda A_\lambda (H_u^+ H_d^- - H_u^0 H_d^0)S + t_S S + \rm{h.c.}\right].
\end{align}
We assume that all the coefficients in the potentials are real so
that any explicit CP violation does not occur other than the
Cabibbo-Kobayashi-Maskawa phase.}
The VEVs of the charged Higgs field should be vanishing in order to
obtain a successful EW symmetry breaking minimum.
One of the VEVs of the charged Higgs field, e.g., $\left\langle
  H_u^+\right\rangle$ can be made zero with positive $\left\langle
  H_u^0\right\rangle=v_u$ by the $SU(2)$ gauge choice. The other one,
however, is not guaranteed to vanish in contrast to the MSSM case in
which the minimization condition $\partial V/\partial H_u^+=0$ gives
$\left\langle H_d^-\right\rangle=0$.
Still, it can be shown that $\left\langle
  H_d^- \right\rangle = 0$ satisfies the extremum conditions and it
will turn out to be local minimum if all masses squared of the charged
Higgs sector are non-negative.
Therefore, it is reasonable to assume that $\left\langle H_d^-
\right\rangle = 0$ and the charged Higgs boson masses squared are required to
be positive. The Higgs potential can then be written as
\begin{eqnarray}
    V_{\rm{Higgs}}&=&\left(m_{H_u}^2+\lambda^2|S|^2\right)|H_u^0|^2+\left(m_{H_d}^2+\lambda^2|S|^2\right)|H_d^0|^2
    -\left[\lambda(A_\lambda S + \mu_S^2) H_u^0 H_d^0 + \rm{h.c.}
    \right]\nonumber\\
    &&+\frac{g_1^2 + g_2^2}{8}\left(|H_u^0|^2-|H_d^0|^2\right)^2+\lambda^2 \left|H_u^0 H_d^0\right|^2 +m_S^2|S|^2+\left[ t_S S + \rm{h.c.}\right]+\mu_S^4.
\end{eqnarray}
Expanding the neutral Higgs fields around their VEVs, one gets
\begin{eqnarray}
  H_u^0 &=& v_u + \frac{H_{uR}+iH_{uI}}{\sqrt{2}}, \\
  H_d^0 &=& v_d + \frac{H_{dR}+iH_{dI}}{\sqrt{2}}, \\
  S &=& v_S + \frac{S_{R}+iS_{I}}{\sqrt{2}}.
\end{eqnarray}
\noindent
In general, the VEVs of the neutral Higgs fields can have non-trivial
phases that induce the spontaneous CP violation. However, as shown in
\cite{Romao:1986}, there cannot be such spontaneous CP violation in
the minimal PQ-NMSSM since the extrema with non-trivial
phases are local maxima rather than minima.
Accordingly, one can always set the VEVs of the neutral Higgs fields to
be real.
The equations of motion $\partial V_{\rm Higgs}/\partial
H_u^0=\partial V_{\rm Higgs}/\partial H_d^0 =\partial V_{\rm
  Higgs}/\partial S=0$ at the vacuum are
{\allowdisplaybreaks \begin{align}
  m_{H_u}^2+\mu_{\rm eff}^2+\lambda^2 v_d^2
  +\frac{g_1^2+g_2^2}{4}(v_u^2-v_d^2)-b_{\rm eff} / \tan\beta &=0,\\
  m_{H_d}^2+\mu_{\rm eff}^2+\lambda^2 v_u^2
  +\frac{g_1^2+g_2^2}{4}(v_d^2-v_u^2)-b_{\rm eff}  \tan\beta &=0,\\
  v_S\left[ m_S^2+\lambda^2 v^2 \right] + t_S  - \lambda A_\lambda v_u
  v_d &=0, \label{Svac}
\end{align}
where $\tan\beta \equiv v_u/v_d$, $v^2 \equiv v_u^2 + v_d^2$,
$\mu_{\rm eff} \equiv \lambda v_S$, and $b_{\rm eff} \equiv \mu_{\rm
  eff} A_\lambda +\lambda \mu_S^2$.}
These three minimization equations can also be cast into the following form,
{\allowdisplaybreaks \begin{align}
% \begin{eqnarray}
  \sin2\beta &=\frac{2b_{\rm eff}}{2\mu_{\rm eff}^2+m_{H_u}^2+m_{H_d}^2+\lambda^2 v^2},\\
  \frac{1}{2} m_Z^2 &=\frac{m_{H_d}^2-m_{H_u}^2 \tan^2\beta}{\tan^2\beta-1}-\mu_{\rm eff}^2,\\
  v_S &=\frac{\lambda A_\lambda v^2 \sin2\beta-2t_S}{2(m_S^2+\lambda^2 v^2)},
% \end{eqnarray}
\end{align}
% with $m_Z^2=g^2v^2$
where $v \simeq 174$ GeV.}
In general, there can be false vacua that do not satisfy the proper EW vacuum
conditions, i.e. $v^2 \simeq (174\,{\rm GeV})^2$, $v_u,\, v_d,\, v_S \neq 0$.
For the EW vacuum to be stable, the false vacua
should not be deeper than the EW vacuum or distant enough to
take a longer time to decay than the age of the Universe.
% Such conditions will impose constraints on the parameter space.
Finding such conditions, however, would be in need of systematic studies as
carried out in \cite{Kanehata:2011ei}, which consider the case of the
$\mathbb{Z}_3$-invariant NMSSM.
Although such extensive works in the case of the PQ-NMSSM are beyond the scope
of this paper, we here leave a comment on the simplest condition for the false vacuum
with $v_u = v_d =0$, $v_S \neq 0$.\footnote{For other false vacua,
numerical analyses would be required even along simple field directions.}
For this false vacuum, the minimum value of the potential is given as
\begin{equation}
  V_{\rm False, min} = -\frac{t_S^2}{m_S^2}+\mu_S^4.
\end{equation}
On the other hand, the potential value at the EW minimum is
\begin{equation}
  V_{\rm True, min} = -\lambda^2 \frac{m_Z^4 \sin^2 2\beta}{4g^4}-
  \frac{m_Z^4 \cos^2 2\beta}{4g^4}+\frac{2\mu_{\rm eff}}{\lambda} t_S
  + \frac{1}{\lambda^2}m_S^2 \mu_{\rm eff}^2 + \mu_S^4.
\end{equation}
with $g^2\equiv (g_1^2+g_2^2)/2$.
And, if one imposes the condition that $V_{\rm True, min} \leq V_{\rm False,
  min}$, it can be shown that this is always satisfied for $|t_S| \gg
|A_\lambda| v^2$.
For $t_S$ around the EW scale or positive value,
$V_{\rm False, min}$ can be much deeper than $V_{\rm True, min}$ depending on the parameter values,
or $m_S^2$ can be negative that it can break the stability of the vacuum.
Therefore, in the following sections for the
phenomenological analysis, we confine the parameter space to the large
$-t_S \sim {\cal O}({\rm TeV})^3$.
The large negative $t_S$ value will make the singlet
scalar heavy enough to be decoupled as can be seen in (\ref{Svac}).
For the small $t_S \lesssim {\cal O}({\rm 100 GeV})^3$ region,
a light singlet-like scalar or pseudoscalar in the Higgs sector still could give an
interesting and different phenomenology \cite{Jeong:2012ma}. However,
more complicated investigation of the parameter space considering the above vacuum
stability conditions should be performed for the region, which is beyond the scope of the present work.

We now expand the Higgs scalar potential around the EW vacuum.
By collecting quadratic terms and eliminating the soft mass terms
through the minimization equations, we get, for the CP-even Higgs
fields in the $(H_{dR},\,H_{uR},\,S_{R})$ basis,
\begin{equation} \label{CPeven}
  {\cal M}_S^2=
  \begin{pmatrix}
    m_Z^2\cos^2\beta+m_A^2\sin^2\beta & (2\lambda^2 v^2-m_A^2-m_Z^2)\sin\beta\cos\beta & \lambda v (2\mu_{\rm eff} \cos\beta - A_\lambda\sin\beta) \\
    & m_Z^2\sin^2\beta+m_A^2\cos^2\beta & \lambda v (2\mu_{\rm eff} \sin\beta - A_\lambda\cos\beta) \\
    & & m_S^2+\lambda^2v^2
  \end{pmatrix},
\end{equation}
where $m_A^2 \equiv 2b_{\rm eff}/\sin 2\beta$.
Upon diagonalizing the matrix, one eventually obtains the following
tree-level lightest Higgs boson mass in the limit of $m_A \gg m_Z$ and
$m_S \gg \mu_{\rm eff}, A_\lambda$.
\begin{equation} \label{eq:Higgs_tree}
  m_{h, \rm tree}^2 \simeq m_Z^2 \cos^2 2\beta+\lambda^2 v^2\left( \sin^2 2\beta
    -\frac{(2\mu_{\rm eff}  - A_\lambda\sin 2\beta)^2}{m_S^2}\right) .
\end{equation}
The second and third terms proportional to $\lambda^2$ come from the
NMSSM Yukawa coupling $\lambda S H_u H_d$ in the superpotential, which
can significantly enhance the tee-level lightest Higgs boson mass
compared to the MSSM.
The second term is due to the doublet scalar quartic coupling
$\lambda^2 \left|H_u^0 H_d^0\right|^2$ and is sensitively becoming
small for the large $\tan\beta$ ($\sin2\beta \sim 2/\tan\beta$), while
the third term is from the doublet-singlet mixing and
$\lambda^2|S|^2|H_{u,d}^0|^2$ terms in the scalar potential.

For the CP-odd Higgs fields,
\begin{equation} \label{CPodd}
  {\cal M}_P^2=
  \begin{pmatrix}
    m_A^2\sin^2\beta & m_A^2 \sin\beta\cos\beta & \lambda v A_\lambda\sin\beta \\
    & m_A^2\cos^2\beta & \lambda v A_\lambda\cos\beta \\
    & & m_S^2+\lambda^2v^2
  \end{pmatrix}
\end{equation}
in the $(H_{dI},\,H_{uI},\,S_{I})$ basis.
After dropping the Goldstone mode obtained by rotating the upper
$2\times2$ matrix, the following mass matrix appears in the
$(A,\,S_I)$ basis,
\begin{equation}
  {\cal M}_{P^\prime}^2=
  \begin{pmatrix}
    m_A^2 & \lambda v A_\lambda\\
    \lambda v A_\lambda & m_S^2+\lambda^2v^2
  \end{pmatrix}.
\end{equation}
The corresponding mass eigenvalues are
\begin{equation}
  m_{A_0, A_1,\rm tree}^2 = \frac{1}{2}\left( m_A^2+m_S^2+\lambda^2 v^2
    \mp \sqrt{(m_S^2-m_A^2+\lambda^2 v^2)^2+4\lambda^2 v^2
      A_\lambda^2}\right).
\end{equation}

For the charged Higgs fields,
\begin{equation}
  {\cal M}_{\pm}^2=
  (m_A^2+m_W^2-\lambda^2 v^2)
  \begin{pmatrix}
    \cos^2\beta & \cos\beta\sin\beta\\
    \cos\beta\sin\beta & \sin^2\beta
  \end{pmatrix}
\end{equation}
in the $(H_u^+,\,H_d^{-\ast})$ basis. By rotating the matrix by the angle
$\pi/2-\beta$, one can obtain the Goldstone mode and a mass eigenstate with the
mass eigenvalue,
\begin{equation}
  m_{H^\pm,\rm tree}^2 = m_A^2 + m_W^2 - \lambda^2 v^2.
\end{equation}

All the above masses will receive important loop corrections (see
Appendix C in \cite{Ellwanger:2009dp}). For the analysis in
Secs.~\ref{sec:Higgs_pheno} and \ref{sec:collider}, we include the
loop contributions as implemented in the \textsc{Nmssmtools
3.1.0}~\cite{Muhlleitner:2003vg}\footnote{
  The current version of the
  \textsc{Nmssmtools} implements only the $\mathbb{Z}_3$-invariant
  NMSSM. We used modified codes for adapting to the case of the
  PQ-NMSSM.}
%
% CBP: I think we can skip the expressions below.
%
% Here, we just note that the SM-like Higgs boson mass acquires the following
% dominant radiative corrections at one-loop level from the top sector and the singlino
% Yukawa interactions \cite{Jeong11} :
% \begin{equation}
%   \delta m_{h, \rm 1-loop}^2 = \frac{3m_t^4}{4\pi^2v^2}\left[\ln\left(
%       \frac{m_{\widetilde{t}}^2}{m_t^2}\right)+\frac{X_t^2}{m_{\widetilde{t}}^2}
%     \left(1-\frac{X_t^2}{12m_{\widetilde{t}}^2} \right)\right]
%   +\frac{\lambda^4 v^2}{4\pi^2}\ln\left(\frac{M_{\rm SUSY}^2}{\mu_{\rm eff}^2}\right)
% \end{equation}
% where the stop mixing parameter $X_t = A_t - \mu_{\rm eff} \cot \beta$.

\subsubsection{Neutralino sector}
\label{sec:Neutralino}

The singlet superfield $S$ in the NMSSM also significantly changes the neutralino sector
compared with the MSSM. The additional singlino field $\widetilde{S}$ mixes with the neutral
Higgsinos $\widetilde{H}_d^0$, $\widetilde{H}_u^0$ and the gauginos $\widetilde{\lambda}_1$,
$\widetilde{\lambda}_2^3$,
producing a symmetric $5\times5$ mass matrix $M_{\widetilde{\chi}^0}$,
\begin{equation} \label{Neutmatrix}
  {\cal M}_{\widetilde{\chi}^0}=
  \begin{pmatrix}
    M_1 & 0 & -g_1v_d/\sqrt{2} & g_1v_u/\sqrt{2} & 0 \\
    & M_2 & g_2v_d/\sqrt{2} & -g_2v_u/\sqrt{2} & 0 \\
    & & 0 & -\mu_{\text{eff}} & -\lambda v_u \\
    & & & 0 & -\lambda v_d \\
    & & & & 0
  \end{pmatrix}
\end{equation}
in the basis $(-i\widetilde{\lambda}_1,\,-i\widetilde{\lambda}_2^3,\,
\widetilde{H}_d^0,\,\widetilde{H}_u^0,\,\widetilde{S})$.
The diagonalization of this mass matrix and resulting mixing matrices
are computed in Appendix~\ref{sec:Neutmixing}.
% Note that there is no SUSY mass term for the singlino $\widetilde{S}$
% in the minimal PQ-NMSSM. It generates the singlino-like lightest
% neutralino whose mass is induced only by mixing, so generally the mass
% is quite smaller than the one in the MSSM.
An important point to note is that there is no SUSY mass term for
the singlino $\widetilde{S}$ in the minimal PQ-NMSSM. The
singlino-like neutralino mass is induced only by mixing, and thus making
the corresponding mass eigenvalue generically quite small. The
lightest neutralino mass appears to be
\begin{eqnarray}
  m_{\widetilde{\chi}_1^0} &=& -2\left(\mu_{\rm eff} N_{13}
    N_{14}+\lambda v \cos\beta N_{14} N_{15}
    +\lambda v \sin \beta N_{13} N_{15}\right)  \nonumber \\
  &&+\sqrt{2} v \left(g_1 N_{11}-g_2 N_{12}\right)\left(-N_{13} \cos\beta + N_{14}
  \sin\beta\right)+M_1 N_{11}^2+M_2 N_{12}^2 \nonumber \\
  &\simeq & \frac{\lambda^2 v^2}{\mu_{\rm eff}}
  \left[ \sin 2\beta - \frac{\lambda^2 v^2}{\mu_{\rm eff}^2}\sin 2\beta
    -\left(\frac{g_1^2v^2}{2\mu_{\rm eff} M_1} +\frac{g_2^2v^2}{2\mu_{\rm eff} M_2}\right)
    \cos^2 2\beta +{\cal O}(\frac{v^4}{\mu_{\rm eff}^4})\right]
  \label{eq:chi1_mass}
\end{eqnarray}
when $M_1, M_2 \sim \mu_{\rm eff} \gg \lambda v$. Here, $N_{1i}$'s
denote the neutralino mixing components given in
Appendix~\ref{sec:Neutmixing}. The larger $\tan \beta$
makes the lightest neutralino mass smaller. Moreover, one can see
that the lightest neutralino becomes lighter as $\mu_{\rm eff}$
increases. It can also be easily checked that the mass becomes
zero as $\mu_{\rm eff}$ vanishes if we consider decoupling of the
gauginos, i.e., $M_1,\,M_2 \gg \mu_{\rm eff},\,\lambda v$. This
observation implies that there must exist a maximum value of the
lightest neutralino mass at a certain value of $\mu_{\rm eff}$ for
a fixed value of $\lambda v$ in decoupling limit of the gauginos.
We can find that the upper bound of the lightest neutralino mass
is given by
\begin{equation}
  m_{\widetilde{\chi}_1^0} = \lambda v \cos \beta \quad{\rm at} ~\mu_{\rm eff} = \lambda v \sin \beta
  \label{eq:max_neut}
\end{equation}
for $\tan \beta > 1$, which gives $m_{\widetilde{\chi}_1^0} \lesssim 85~{\rm GeV}$ for
$\lambda = 0.7$.

%%%%%%%%%%%%%%%%%%%%%%%%%%%%%%%%%%%
%%%%% by kjb %%%%%%%%%%%%%%%%%%%%%
%%%%%%%%%%%%%%%%%%%%%%%%%%%%%%%%%

\section{Higgs phenomenology of the PQ-NMSSM}
\label{sec:Higgs_pheno}

In this section, we discuss the Higgs phenomenology of the
PQ-NMSSM in the light of the recent ATLAS and CMS discovery of a
125 GeV Higgs boson. We will show viable parameter spaces that can
give the presumed Higgs boson mass without conflicting any
existing phenomenological constraints, and discuss their
feasibility from the point of view of naturalness.

As discussed in Sec.~\ref{sec:Higgs}, the relevant parameters for the
Higgs sector are $\lambda$, $\tan \beta$, $\mu_{\rm eff}$, $m_A$ (or
$A_\lambda$), $\mu_S^2$, and $t_S$.
For $\mu_S^2$, we take a weak scale value of ${\cal O}(100\,{\rm
  GeV})^2$ since it affects the Higgs sector only via the $b_{\rm
  eff}$ $(= \mu_{\rm eff} A_\lambda +\lambda \mu_S^2)$ term so that
the variations of $\mu_{\rm eff}$ or $m_A$($A_\lambda$) include its
effect. For $m_A$ and $t_S$, we select their proper values that
can give $m_S^2 \gg m_A^2 \gg m_Z^2$ in order to make the tree-level
Higgs boson mass as large as possible according to
(\ref{eq:Higgs_tree}).\footnote{
  The region of $m_A^2 \sim m_Z^2$ will also be considered in the
  subsequent discussion with regard to the suppression of the ratio of
  the Higgs invisible decay. However, it will turn out to be
  unfavorable to the constraint on the charged Higgs boson.}
A large value of $\lambda$ is needed to increase the Higgs boson mass
through the specific contribution of the NMSSM.
%
% But we should be careful on perturbativity breaking at some high
% scale by large $\lambda$ driven by renormalization group running.
% It can be shown that the perturbativity breaking scale rapidly
% decreases as $\tan \beta$ becomes small.
% In Table~\ref{table:landau}, we show the perturbativity breaking
% scales in the small $\tan \beta$ region for a fixed value of $\lambda
% = 0.7$.
% If the perturbativity breaking scale is lower than the PQ scale
% of $10^9 - 10^{12}$ GeV, the PQ-NMSSM cannot make sense.
% As shown in Table \ref{table:landau}, $\lambda = 0.7$ is a quite marginal
% choice for the PQ-NMSSM to be viable in small $\tan \beta$ region.
% Therefore, we will choose 0.7 for $\lambda$.
%
However, an investigation into the scale of the perturbativity
breaking by $\lambda$ driven by renormalization group running
should be preceded to avoid the scale being lower than the PQ
scale of $10^9 - 10^{12}$ GeV. It can be shown that the
perturbativity breaking scale decreases rapidly as $\tan \beta$
becomes small. In Table~\ref{table:landau}, we show the
perturbativity breaking scales in the small $\tan \beta$ region
for a fixed value of $\lambda = 0.7$, which is a marginal choice
for the PQ-NMSSM to be viable in the small $\tan\beta$ region.

\begin{table}[h!]
  \caption{Perturbativity breaking scale $\Lambda$ for $\tan\beta\lesssim2$.}
  \label{table:landau}
  \centering
  % \begin{tabular}{|c|c|c|c|c|c|c|c|}
  \begin{tabular}{c|ccccccc}
    \hline\hline&&&&&&&\\[-2mm]
    $\tan\beta$ & 1.1 & 1.2 & 1.3 & 1.4 & 1.5 & 1.7 & 1.9
    \\[2mm]
    \hline&&&&&&&\\[-2mm]
    $\Lambda$ (GeV) & $10^{10}$ & $10^{11}$ & $10^{12}$ & $10^{13}$
    & $10^{14}$ & $10^{15}$ & $10^{16}$
    \\[2mm]
    \hline\hline
  \end{tabular}
\end{table}

One of the most important features of the PQ-NMSSM is that the lightest
neutralino is relatively lighter than that of the MSSM. Such a light
neutralino can raise conflicts with several phenomenological
constraints. Furthermore, since the neutralino sector shares some parameters
with the Higgs sector, the constraints can also place serious
restrictions on the viable parameter space for the 125 GeV Higgs signals.
% So we need to be careful on the neutralino sector while investigating
% the parameter space.
As for the neutralino sector, the relevant parameters are
$\lambda$, $\tan \beta$, $\mu_{\rm eff}$, $M_1$, and $M_2$ as can be seen
in Sec.~\ref{sec:Neutralino}.
% The neutralino sector shares the parameters
% $\lambda$, $\tan \beta$, $\mu_{\rm eff}$ with the Higgs sector.
Among them, the overlapping parameters with the Higgs sector are
$\lambda$, $\tan\beta$, and $\mu_{\rm eff}$. The phenomenological
constraints on the light neutralino come from
\begin{enumerate}[I.]
\item $Z$ invisible decay ($Z \rightarrow \widetilde{\chi}_1^0 \widetilde{\chi}_1^0$),
\item $\widetilde{\chi}_1^0 \widetilde{\chi}_2^0$ production at LEP II
  ($e^+ e^- \rightarrow \widetilde{\chi}_1^0 \widetilde{\chi}_2^0$),
\item Higgs invisible decay ($h \rightarrow \widetilde{\chi}_1^0 \widetilde{\chi}_1^0$).
\end{enumerate}
% The constraints I and II will be discussed in
% Sec.~\ref{sec:neut_const}, and III will be discussed in
% Sec.~\ref{sec:Higgs_inv}.
These constraints impose additional restrictions on the parameter
space of $\tan \beta$ and $\mu_{\rm eff}$. In particular, the
constraints I and III motivate us to consider two scenarios according
to the value of $\tan \beta$.
For small $\tan \beta \lesssim 1.6$, the lightest neutralino turns out to
be heavier than the half of the 125 GeV Higgs mass so that the
constraints I and III are satisfied by kinematics. For large $\tan
\beta \gtrsim 1.6$, however, it should be considered a non-trivial mechanism
of suppressing the Higgs invisible decay.
% Sec.~\ref{sec:small_tan} will be devoted to the small $\tan
% \beta$ scenario, while the large $\tan\beta$ scenario will be
% considered in Sec.~\ref{sec:large_tan}.
%
% For the remaining parameters $M_1$ and $M_2$ in the neutralino sector,
% we choose two different gaugino mass relations for the following reasons.
% Firstly, we will not consider too heavy gluino that cannot make any
% eligible new physics signature in the LHC experiment. %at 8 TeV run of LHC.
% But, at the same time, the gluino should be heavy enough to be above the
% region ruled out by the current search result of the LHC.
% Therefore, we consider 1 TeV gluino under the circumstances of heavy
% enough first two generation squark masses~\cite{atlas:susy}.
% The EW gaugino masses are determined by the gluino mass with some
% proportionality assuming specific scenarios.
% We consider two possibilities for the EW gauginos to be light and heavy.
% The first one is the GUT relation of the
% gaugino masses, and the other one is the unified gaugino mass in the
% TeV scale mirage mediation~\cite{Choi:2007ka}.
% That is,

For our study with  specific choices of the usual MSSM parameters,
the gluino mass parameter $M_3$ is taken to be 1 TeV, which is
above the region ruled out by current search
results~\cite{Chatrchyan:2012mf}, while being able to be accessed in
the LHC experiment. For the same purpose, we set the first two
generation squarks at $\sim 2$ TeV. In regard to the remaining
gaugino mass parameters, $M_1$ and $M_2$, we consider two kinds of
scenarios, the grand unification theory (GUT) relation of the
gaugino masses and the unified gaugino masses in the TeV scale
mirage mediation~\cite{Choi:2005uz},
\begin{align}
  \text{GUT}:&\quad 6M_1=3M_2=M_3=1\text{ TeV},\\
  \text{TeV mirage}:&\quad M_1=M_2=M_3=1\text{ TeV}.
\end{align}
In the case of the GUT relation,
the lightest neutralino becomes generically lighter than in the case
of decoupled heavy EW gauginos since the light EW gauginos can have a
mixing with the Higgsino and the singlino.
Furthermore, the mixing can deliver important phenomenological
consequences through affecting the coupling of Higgs to the lightest
neutralino pair as to be discussed in the following subsections.
On the other hand, in the case of the TeV mirage, the EW gauginos are
heavy enough so that the relation (\ref{eq:max_neut}) holds to give
heavier lightest neutralino, and the lightest neutralino is almost
composed of Higgsino and singlino.\footnote{Here we concentrate on the
minimal PQ-NMSSM. Phenomenological implication of  the TeV-scale
mirage mediation in the singlet extension of the MSSM (\ref{S-NMSSM})
was discussed in \cite{Asano:2012sv}.}

In Subsecs.~\ref{sec:neut_const} and \ref{sec:Higgs_inv}, we
review the constraints I, II, and III in the case of the light
neutralino. We then proceed to analyze two scenarios depending on
$\tan \beta$ in Subsecs.~\ref{sec:small_tan} and
\ref{sec:large_tan}.

\subsection{LEP II constraints on the light neutralino}
\label{sec:neut_const}

\begin{figure}[t!]
  \begin{center}
    \includegraphics[width=5cm]{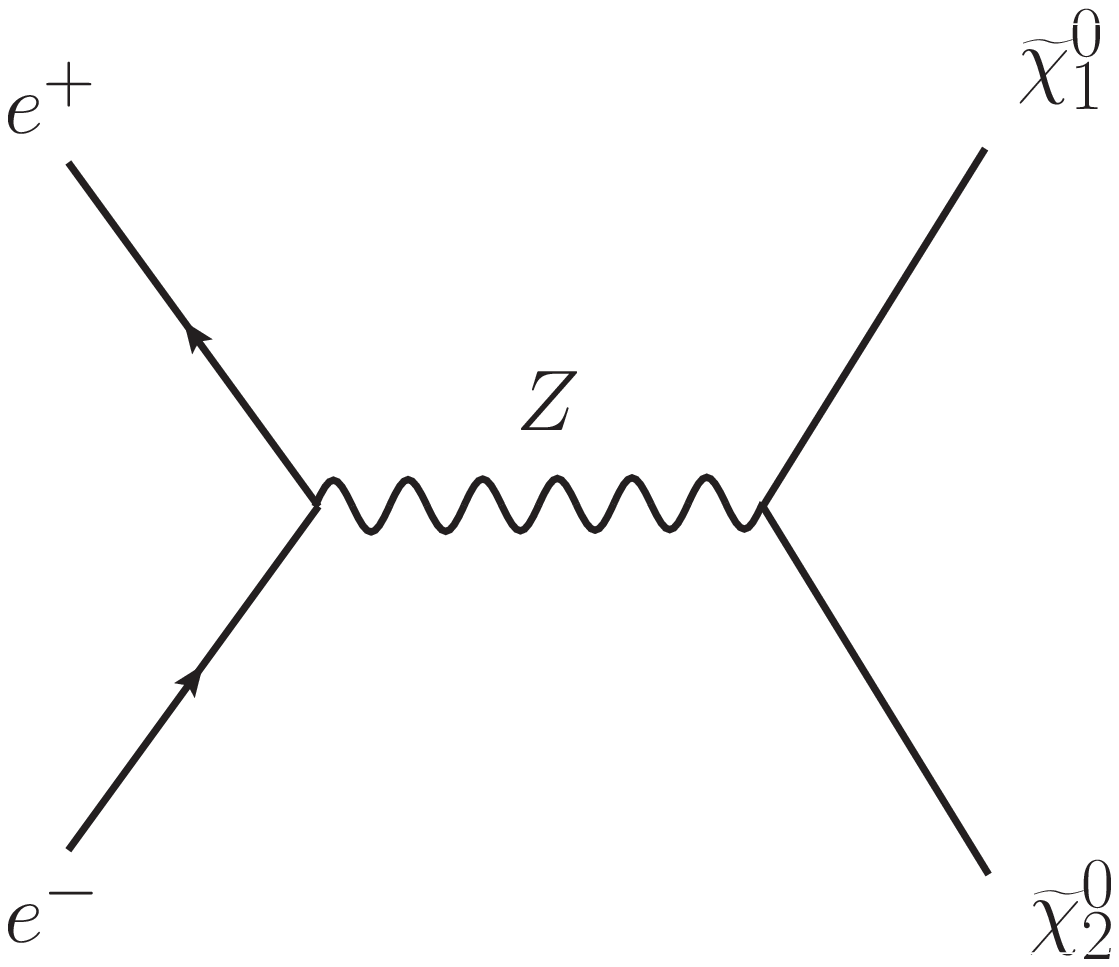}~~~~~
    \includegraphics[width=4.1cm]{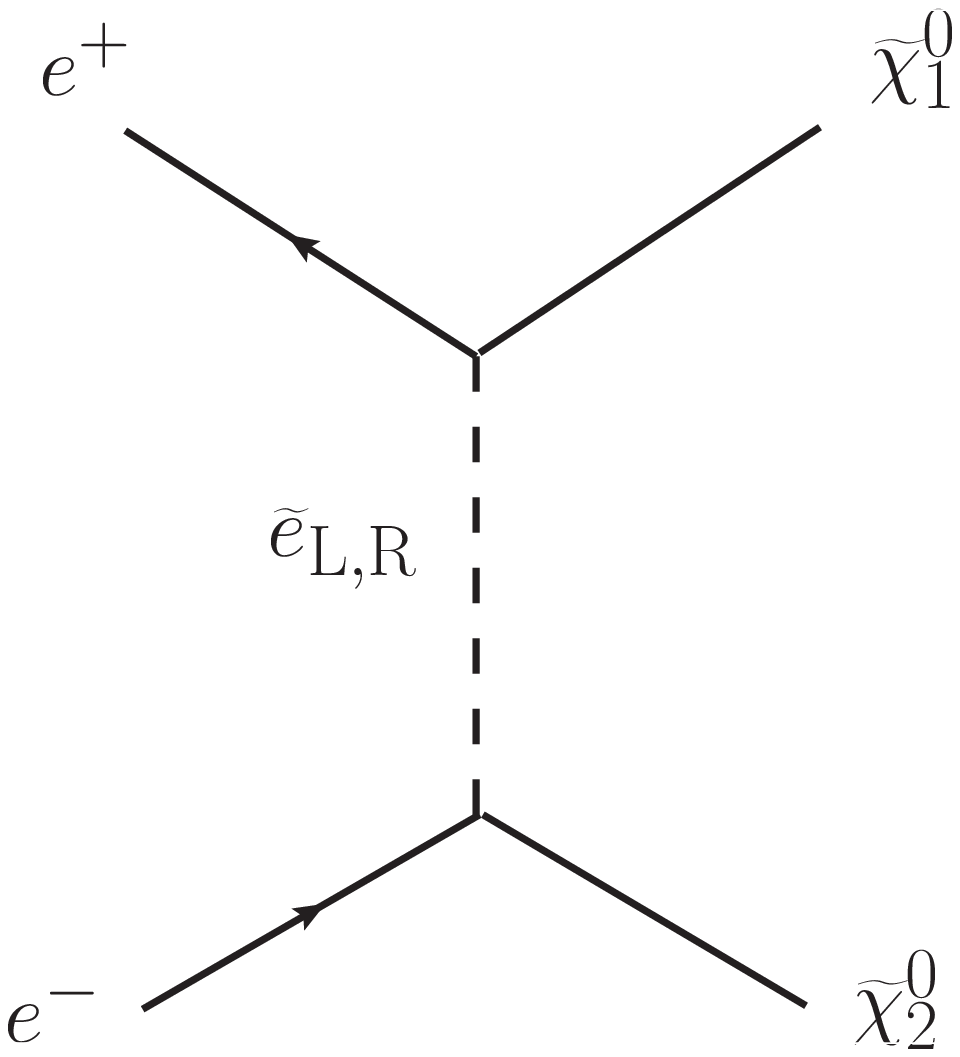}~~~~~
    \includegraphics[width=4.1cm]{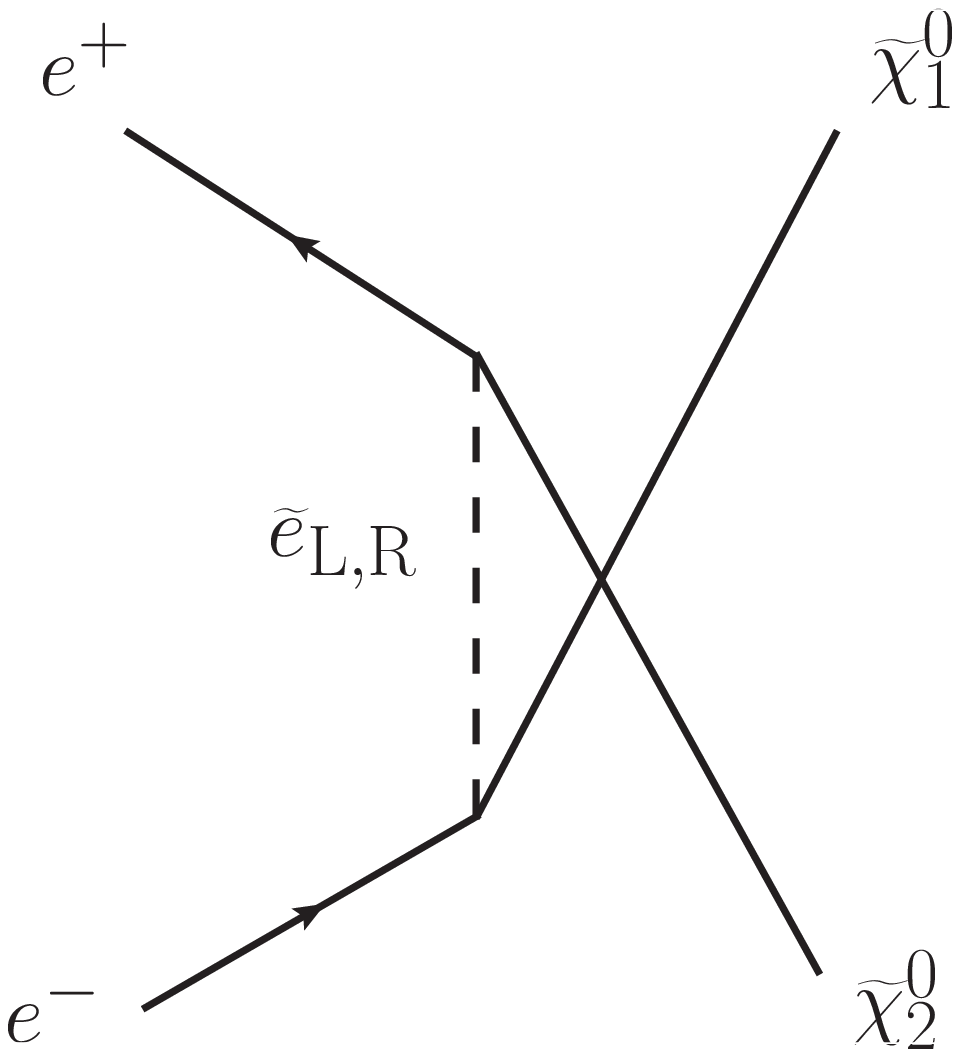}
    \caption{\label{fig:FeynNeutprod}Feynman diagrams for the neutralino production
      $e^{+}e^{-}\to\widetilde{\chi}^0_1\widetilde{\chi}^0_2$}
  \end{center}
\end{figure}

In the PQ-NMSSM, not only the lightest singlino-like neutralino
$\widetilde\chi_1^0$ but also the next-to-lightest neutralino
$\widetilde\chi_2^0$ can be light enough to get constrained by the $Z$ invisible decay and/or
the direct production in the $e^+ e^-$ scattering of the LEP II
experiment through the processes shown in Fig.~\ref{fig:FeynNeutprod}.

The $Z$ boson decay rate to a lightest-neutralino pair is given by
\begin{equation}
  \Gamma(Z\to\widetilde{\chi}_1^0 \widetilde{\chi}_1^0)=\frac{g_2^2}{4\pi}\frac{(N_{13}^2-N_{14}^2)^2}{24\cos^2\theta_W}m_Z\biggl[1-\biggl(\frac{2m_{\widetilde{\chi}_1^0}}{m_Z}\biggr)^2\biggr]^{3/2},
\end{equation}
where $N_{1i}$'s are neutralino mixing components shown explicitly in
Appendix~\ref{sec:Neutmixing}.
From the constraint on the $Z$ invisible decay,
$\Gamma_{\text{inv}}<3$ MeV~\cite{Nakamura:2010zzi}, we find that
\begin{equation}\label{Zinvbound}
  |N_{13}^2-N_{14}^2|\lesssim0.13.
\end{equation}

In the LEP II experiment, the next-to-lightest neutralino can be
produced in association with the lightest neutralino as in
Fig.~\ref{fig:FeynNeutprod}.
% For the center-of-mass energy $\sqrt{s}=208$ GeV of LEP II, we can
% neglect the selectron exchange diagrams by the assumption that
% selectrons are much heavier than $Z$ boson.
If one assumes that the sleptons are much heavier than the $Z$ boson,
the slepton exchange diagrams can be neglected in the LEP II experiment.
Then, the production cross section is given as
\begin{equation}
  \begin{split}
    \sigma(&e^+e^-\to Z^\ast\to\widetilde{\chi}_1^0\widetilde{\chi}_2^0) \\
    =&\frac{1}{96\pi}\frac{g_2^2}{\cos^4\theta_W}\biggl(\frac14-\sin^2\theta_W+2\sin^4\theta_W\biggr)(N_{13}N_{23}-N_{14}N_{24})^2\frac{s}{(s-m_Z^2)^2}\\
    &\times\biggl[1-\frac{2(m_{\widetilde{\chi}^0_1}^2+m_{\widetilde{\chi}^0_2}^2)}{s}+\frac{(m_{\widetilde{\chi}^0_2}^2-m_{\widetilde{\chi}^0_1}^2)^2}{s^2}\biggr]^{1/2}\\
    &\times\biggl[1-\frac{6m_{\widetilde{\chi}^0_1}m_{\widetilde{\chi}^0_2}
      +(m_{\widetilde{\chi}^0_1}^2+m_{\widetilde{\chi}^0_2}^2)}{2s}-\frac{(m_{\widetilde{\chi}^0_2}^2-m_{\widetilde{\chi}^0_1}^2)^2}{2s^2}\biggr].\label{eq:eetochichi}
  \end{split}
\end{equation}
The production cross section is given upper bounds depending on the masses
of the neutralinos by the OPAL analysis result~\cite{Abbiendi:2003sc}.
In the case that the lightest neutralino is nearly massless, the most
conservative bound is given as follows.
\begin{equation}
  \sigma(e^+e^-\to\widetilde{\chi}^0_1\widetilde{\chi}^0_2) < 10\,{\rm fb}
\end{equation}
with Br$(\widetilde{\chi}^0_2\to Z\widetilde{\chi}^0_1)=1$.
However, such an upper limit of the production cross section can be
moderated in the case that the lightest neutralino is heavier.
This will be discussed in more detail in Subsec.~\ref{sec:small_tan}.

\subsection{Higgs invisible decay}
\label{sec:Higgs_inv}

The light neutralino states can also induce the invisible decay of
the Higgs boson. If there exists substantial invisible decay ratio
of the Higgs boson, the visible decay modes such
$h\to\gamma\gamma$ or $h\to WW$/$ZZ$ will be significantly reduced. As
this would make it difficult to explain the recent discovery of
the 125 GeV Higgs boson signal in the ATLAS and CMS experiments,
it is of particular importance to consider the Higgs invisible
decay in the plausible parameter space.

The invisible decay of the lightest CP-even Higgs boson is determined
by the Higgs-neutralino coupling shown in Appendix~\ref{sec:Higgs-neut
  coup.},
\begin{equation} \label{coupl:Hchichi}
  g_{h \widetilde{\chi}_i^0\widetilde{\chi}_j^0}=\frac{\lambda}{\sqrt{2}}(S_{11}\Pi_{ij}^{45}+S_{12}\Pi_{ij}^{35}+S_{13}\Pi_{ij}^{34})
  +\frac{g_1}{2}(S_{11}\Pi_{ij}^{13}-S_{12}\Pi_{ij}^{14})-\frac{g_2}{2}(S_{11}\Pi_{ij}^{23}-S_{12}\Pi_{ij}^{24}),
\end{equation}
where $\Pi_{ij}^{ab} \equiv N_{ia}N_{jb}+N_{ib}N_{ja}$.
Here, $S_{ab}$ and $N_{ij}$ are the CP-even Higgs mixing matrix and
the neutralino mixing matrix, respectively.
% , which are defined in the
% appendix.
% We need to compute the neutralino mixing components to find the
% condition for vanishing coupling.
As analyzed in Appendix~\ref{sec:Neutmixing} including the first,
second, and dominant third-order contributions, the  mixing
components in the neutralino mass matrix are
 {\allowdisplaybreaks
\begin{align}
  N_{11} \simeq &-\frac{g_1\lambda v^2 \cos2\beta}{\sqrt{2}M_1\mu_{\text{eff}}}\label{N_11},\\
  N_{12} \simeq &\frac{g_2\lambda v^2 \cos2\beta}{\sqrt{2}M_2\mu_{\text{eff}}}\label{N_12},\\
  N_{13} \simeq &-\frac{\lambda
    v\cos\beta}{\mu_{\text{eff}}}-\frac{\lambda
    v^3}{2\mu_{\text{eff}}^2}\biggl(\frac{g_1^2}{M_1}+\frac{g_2^2}{M_2}\biggr)\cos2\beta\sin\beta+\frac{3\lambda^3v^3}{\sqrt{2}\mu_{\text{eff}}^3}\sin2\beta\sin\beta \label{N_13},\\
  N_{14} \simeq &-\frac{\lambda v\sin\beta}{\mu_{\text{eff}}}\label{N_14},\\
  N_{15} \simeq & 1\label{N_15}.
\end{align}
We here include only the leading term for each component except
$N_{13}$, for which the second and third terms in (\ref{N_13}) can
be sizable when $\tan\beta\gtrsim(\mu/\lambda v)^2$.}
By putting these terms into (\ref{coupl:Hchichi}), one can find
$g_{h\widetilde{\chi}_1^0\widetilde{\chi}_1^0}$ for the coupling
between the lightest CP-even Higgs and a lightest-neutralino pair.
In the limiting case of vanishing doublet-singlet mixing in the
Higgs sector, i.e., $S_{13}=0$, the coupling is given by
\begin{equation}
  \begin{split}
    \frac{g_{h\widetilde{\chi}_1^0\widetilde{\chi}_1^0}}{S_{12}}\simeq
    &-\frac{\sqrt{2}\lambda^2v}{\mu_{\text{eff}}}\biggl(1+c\frac{3m_W^2}{2M_2\mu^3}\biggr)\biggl(1-\frac{3\sqrt{2}\lambda^2v^2}{\mu_{\text{eff}}^2}\biggr)\frac{1}{\tan\beta}\\
    &-\frac{\sqrt{2}\lambda^2v}{\mu_{\text{eff}}}\biggl(c-c\frac{9m_W^4}{4M_2^2\mu_{\text{eff}}^2}-\frac{3m_W^2}{M_2\mu_{\text{eff}}}\biggr),
  \label{coupl:Hchichi2}
  \end{split}
\end{equation}
where $c \equiv S_{11}/S_{12}$ and  $g_2 \simeq 2g_1$. It was
assumed that the GUT relation for the gaugino masses, $M_2 =
2M_1$, and applied a crude approximation, $\sin\beta \simeq 1$,
$\cos\beta \simeq 1/\tan\beta$, $\sin2\beta \simeq 2/\tan\beta$
and $\cos2\beta\simeq 1$ for $\tan\beta\gtrsim3$ to derive the
above relation. Note that the coefficient $c$ corresponds to
$-\tan\alpha$ ($-\pi/2\leq\alpha\leq0$) in the MSSM limit, where
$\alpha$ is the CP-even Higgs mixing angle in the MSSM. For making
this coupling vanishing, we find a relation,
% \begin{equation}
%   \begin{split}
{\allowdisplaybreaks \begin{align}
    \frac{1}{\tan\beta}\simeq&-\biggl(c-c\frac{9m_W^4}{4M_2^2\mu_{\text{eff}}^2}-\frac{3m_W^2}{M_2\mu_{\text{eff}}}\biggr)
    \biggl(1-c\frac{3m_W^2}{2M_2\mu_{\text{eff}}}\biggr)\biggl(1+\frac{3\sqrt{2}\lambda^2v^2}{\mu_{\text{eff}}^2}\biggr)
    \nonumber\\
    =&\biggl[\frac{3m_W^2}{M_2\mu_{\text{eff}}}-c\biggl(1+\frac{9m_W^4}{4M_2^2\mu_{\text{eff}}^2}\biggr)-c^2\frac{3m_W^2}{2M_2\mu_{\text{eff}}}\biggl(1-\frac{9m_W^4}{4M_2^2\mu_{\text{eff}}^2}\biggr)\biggr]\biggl(1+\frac{3\sqrt{2}\lambda^2v^2}{\mu_{\text{eff}}^2}\biggr).
    \label{eq:g_Hchichi=0}
\end{align}
%   \end{split}
% \end{equation}
For $M_2\mu_{\text{eff}}\gg m_W^2$, the coefficients of $c$ and $c^2$
are always positive, so the larger $c$ requires the larger value of
$\tan\beta$.
The effect of the Higgs mixing $c$ will be discussed in
Subsec.~\ref{sec:large_tan}.}

\subsection{\boldmath Phenomenology depending on $\tan\beta$}

\subsubsection{\boldmath $1\lesssim\tan\beta\lesssim2$: heavy
  neutralino scenario}
\label{sec:small_tan}

We now discuss the possibility for the small $1 \lesssim \tan\beta
\lesssim 2$ region. Such small $\tan\beta$ makes
singlino-Higgsino mixing large and thus the lightest neutralino
heavy.
% Then, one obtains the lightest neutralino mass larger than $m_h/2$ so that
% the invisible decays of the $Z$ boson and the Higgs boson are
% kinematically forbidden.
% In this case, however, the next-to-lightest neutralino mass can be somewhat smaller so that the $\widetilde{\chi}_1^0\widetilde{\chi}_2^0$ production might constrain this possibility.
% As discussed in Sec.~\ref{sec:neut_const}, the most conservative bound for such neutralino pair production cross section is 10 fb.
% However, the constraint from the OPAL~\cite{Abbiendi:2003sc} is not a single number but it varies depending on the $m_{\widetilde{\chi}_1^0}$ and $m_{\widetilde{\chi}_2^0}$.
% As can be seen in Fig. \ref{fig:OPAL}, the upper bound becomes weaker
% to 70 fb in the parameter space of our interest,
% i.e. $m_{\widetilde{\chi}_1^0}\gtrsim m_h/2\sim63$ GeV and $120$
% GeV$\lesssim m_{\widetilde{\chi}_2^0}\lesssim 140$ GeV.
%
One can obtain the mass of the lightest neutralino larger than
$m_h/2$ in order to forbid kinematically the invisible decays of
the $Z$ boson and the Higgs boson. In this case, however, it may
concern the $\widetilde{\chi}_1^0 \widetilde{\chi}_2^0$ pair
production as the next-to-lightest neutralino can also be light
enough. As discussed in Sec.~\ref{sec:neut_const}, the most
conservative bound for such a neutralino-pair production cross
section was set by the OPAL result. The upper bound of the cross
section varies according to mass values of the neutralinos. It
becomes as much as 70 fb in the parameter space of our interest,
$m_{\widetilde\chi_1^0} \gtrsim m_h / 2 \sim 63$ GeV and $120
\lesssim m_{\widetilde\chi_2^0} \lesssim 140$ GeV.

\begin{figure}[t!]
  \begin{center}
    \includegraphics[width=10cm]{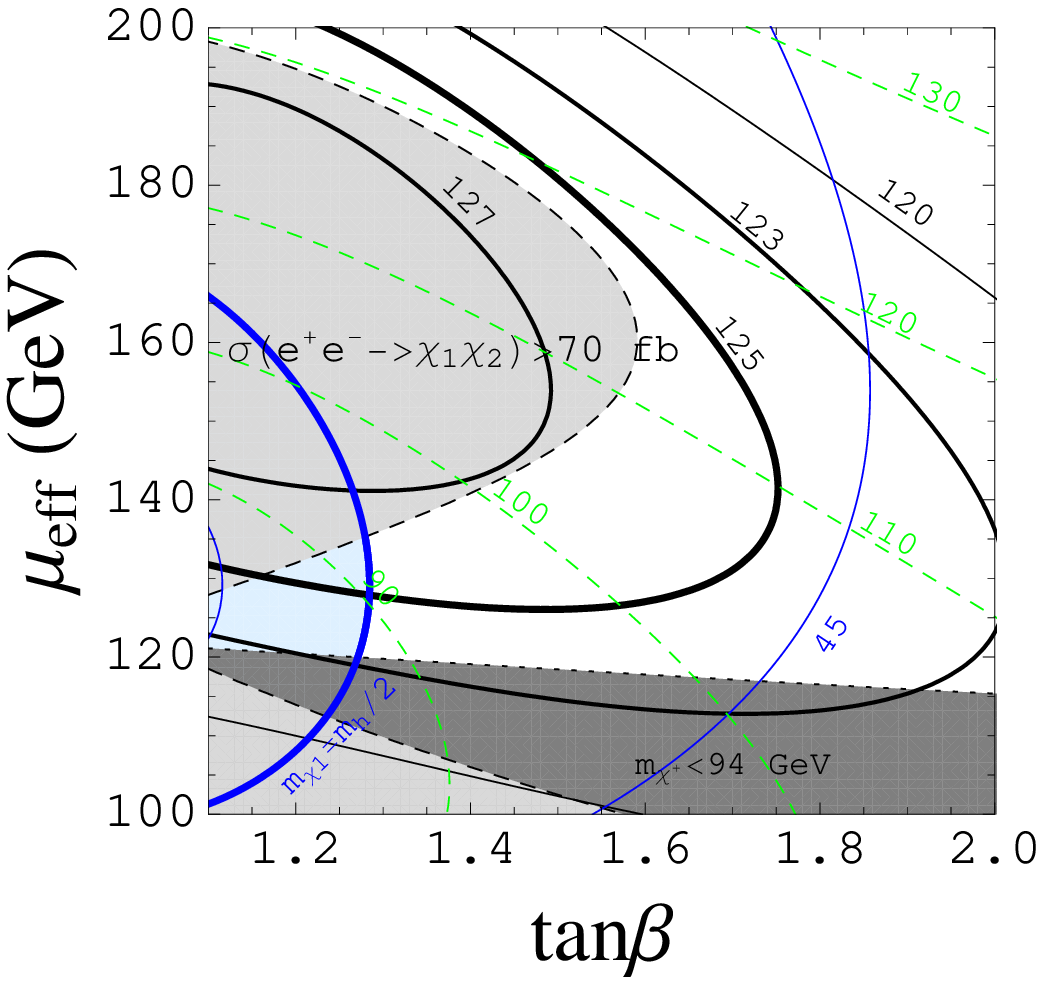}
    \caption{Plot for small $\tan\beta$. Here we set $6M_1=3M_2=M_3=1$
      TeV,  $m_A=350$ GeV,
      $m_{\widetilde{Q}_3}=m_{\widetilde{t}^c}=500$ GeV, $A_t=0$, and
      $\xi_S=-7\times10^7\text{ GeV}^3$. Black curves denote Higgs
      mass in GeV, blue curves denote $m_{\widetilde\chi_1^0}$.
      in GeV, and green dashed curves denote $m_{\widetilde\chi_2^0}$.
      The gray-shaded region is excluded by the OPAL~\cite{Abbiendi:2003sc}.
      \label{fig:lowtan_gut}}
  \end{center}
\end{figure}

\begin{figure}[t!]
  \begin{center}
    \includegraphics[width=10cm]{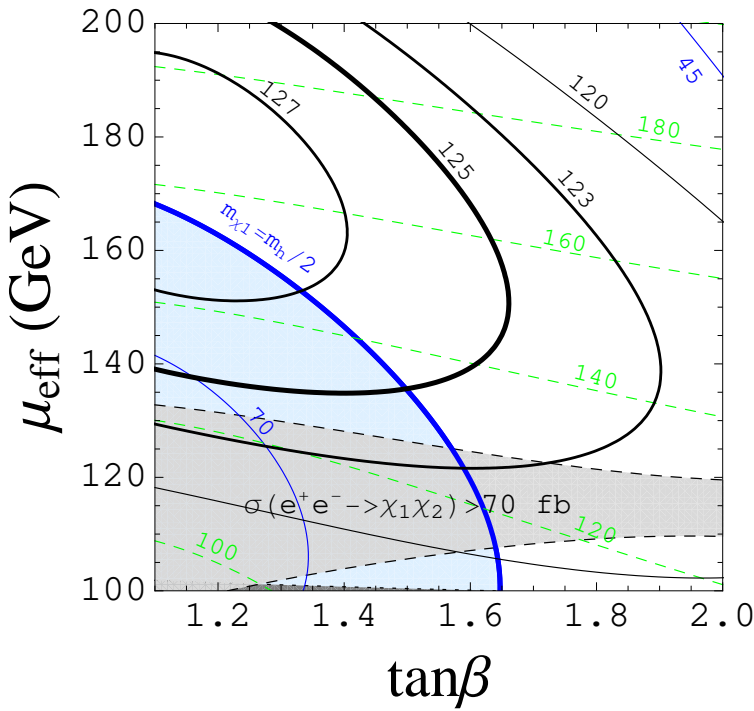}
    \caption{Plot for small $\tan\beta$. Here we set $M_1=M_2=M_3=1$
      TeV,  $m_A=350$ GeV,
      $m_{\widetilde{Q}_3}=m_{\widetilde{t}^c}=500$ GeV, $A_t=0$, and
      $\xi_S=-7\times10^7\text{ GeV}^3$. Black curves denote Higgs
      mass in GeV, blue curves denote $m_{\widetilde\chi_1^0}$.
      in GeV, and green dashed curves denote $m_{\widetilde\chi_2^0}$.
      The gray-shaded region is excluded by the OPAL~\cite{Abbiendi:2003sc}.
      \label{fig:lowtan_mirage2}}
  \end{center}
\end{figure}

The numerical results for the GUT and TeV mirage relations of gaugino
masses are shown in Figs.~\ref{fig:lowtan_gut} and
\ref{fig:lowtan_mirage2}, respectively.
In the case of the GUT relation, light EW gauginos are mixed with
the singlino, and the lightest neutralino becomes lighter.
Moreover, the next-to-lightest neutralino also becomes lighter due to
the Higgsino-gaugino mixing, which increases the
$\widetilde{\chi}_1^0\widetilde{\chi}_2^0$ production in the LEP II
experiment.
Hence, only small region of the parameter space,
$\tan\beta\lesssim1.2$ and $120\,{\rm GeV} \lesssim \mu_{\rm eff}
\lesssim 130$ GeV, is allowed.
On the other hand, in the case of the TeV mirage, the EW gauginos are
heavy so that the lightest and the next-to-lightest neutralinos are mostly composed
of the singlino and the Higgsino. Consequently, the masses of the
lightest two neutralinos are heavier than those of the GUT relation
case, and thereby avoiding the LEP II constraint for the
$\widetilde{\chi}_1^0\widetilde{\chi}_2^0$ production for broader
parameter region, $\tan\beta\lesssim1.5$ and $130\,{\rm GeV} \lesssim
\mu_{\rm eff} \lesssim 170$ GeV.

This small $\tan\beta$ scenario has excellent features in the
naturalness point of view.
The large loop corrections by the stops are not necessary to raise the
Higgs boson mass up to 125 GeV as the tree-level Higgs boson mass
can be raised enough by the $\lambda$-proportional contribution by
virtue of the small $\tan\beta$.
This ameliorates the fine-tuning problem from the stop sector.
Moreover, all massive soft parameters related to the EW symmetry breaking are of order of 100 GeV.
When $\tan\beta=1.3$ and $\mu_{\text{eff}}=135\text{ GeV}$ in Fig. \ref{fig:lowtan_mirage2}, for example, the masses related to the EW symmetry breaking are given by
\begin{equation}
  m_{H_u}\sim m_{H_d}\sim 200\text{ GeV},\quad m_S\sim 600\text{ GeV}, \quad m_{\widetilde{t}}\sim 500{\text{ GeV}}.
\end{equation}
Therefore, the EW symmetry breaking condition can be satisfied
up to 5\% parameter tuning.
% The numerical result is shown in Fig. \ref{fig:lowtan}. In the region of $\tan\beta\lesssim1.5$ and $130$ GeV$\lesssim\mu\lesssim160$ GeV, we can obtain heavy neutralino with mass larger than $m_h/2$ so that LEP constraints can be evaded.
% In this region, $m_A$ and $m_S$ are large enough, so the lightest Higgs is almost SM-like.
% Then, the partial decay width for each channel (i.e. branching fractions) is almost the same as that of SM.
% (The only one exception is $h\to gg$ since the light stop. It is about 0.06 while SM value is about 0.08. Even if stop mass is of order of TeV, this difference still exists. I think such difference comes from the next order calculation of HDECAY code. Hence I don't think we have to concentrate this discrepancy.)


Concerning the cosmology in this scenario, if the LSP is the lightest
neutralino, the substantial Higgsino mixing in the LSP becomes the
cause of the similar cosmological features as the Higgsino dark matter
of the MSSM.
Still, due to the large Higgsino-singlino-Higgs coupling from the
$\lambda S H_u H_d$ term in the superpotential, the direct
detection cross section can be rather large. On the other hand, it
is expected that the amount of missing energy in the LHC
experiment will not be much different from that of the MSSM since
the $m_{\widetilde\chi_1^0}$ is slightly larger than $m_h/2$. The
existence of the light third-generation squarks will lead to the
top or bottom-rich signal. The dark matter and the collider
signatures at the LHC for the small $\tan\beta$ region will be
discussed in more detail in Secs.~\ref{sec:DM} and
\ref{sec:collider}, respectively.

\subsubsection{\boldmath Large $\tan\beta$: very light neutralino scenario}
\label{sec:large_tan}

For large values of $\tan\beta \gtrsim 1.6$, the lightest neutralino
mass becomes smaller than $m_h/2$, and the Higgs
invisible decay mode to a lightest neutralino pair is open. In this
region, it is necessary to examine conditions for suppressing
the Higgs invisible decay.
As argued in (\ref{eq:g_Hchichi=0}), it is plausible to put the Higgs
mixing parameter $c$ as small as possible in order to make the Higgs
invisible decay rate vanishing, while keeping relatively small
$\tan\beta$ by which the quartic coupling $\lambda$ can substantially
raise the tree-level Higgs boson mass.
As an extreme example, we consider vanishing Higgs mixing parameter $c$.
To achieve this for relatively small $\tan\beta$, we set ${\cal
  M}_{S,12}^2=0$ implying
\begin{equation}
  % {\cal M}_{S,12}^2=
  (2\lambda^2v^2-m_A^2-m_Z^2)\cos\beta\sin\beta=0,
  \label{eq:zeromix}
\end{equation}
that is,
\begin{equation} \label{eq:mA}
  m_A^2=2\lambda^2v^2-m_Z^2\approx (146 \text{ GeV})^2.
\end{equation}
In this case, the lightest Higgs boson is mostly the up-type, $H_u$,
for $\tan\beta\gtrsim2$. Here, $m_A^2$ is not the physical CP-odd
Higgs boson mass, but a model parameter defined as $m_A^2 \equiv
2b_{\text{eff}}/\sin2\beta$, which indeed corresponds to the CP-odd
Higgs boson mass in the MSSM limit.
Now, by putting $c=0$ in (\ref{eq:g_Hchichi=0}), one can find
\begin{equation}
  \frac{1}{\tan\beta}\simeq\frac{3m_W^2}{M_2\mu_{\text{eff}}}\biggl(1+\frac{3\sqrt{2}\lambda^2v^2}{\mu_{\text{eff}}^2}\biggr),
\end{equation}
and thus $\tan\beta\approx3$ for $\mu_{\text{eff}}=M_2=300$ GeV.
Although one can suppress the Higgs invisible decay  with
$\tan\beta\approx 3$ in this way, the small $m_A^2$ in (\ref{eq:mA}) can
lead to the light charged Higgs boson.
At tree level, the charged Higgs boson mass is given by
\begin{equation}
  m_{H^{\pm}}^2=\frac{2(\mu_{\text{eff}}B_{\text{eff}}+\hat{m}_3^2)}{\sin2\beta}
  +v^2\biggl(\frac{g_2^2}{2}-\lambda^2\biggr)=m_A^2+m_W^2-\lambda^2v^2,
\end{equation}
leading to $m_{H^\pm} \approx 113$ GeV for the parameter choice (\ref{eq:mA}).
% Such light charged Higgs boson with the mass smaller than 150 GeV and
% $\tan\beta\lesssim4$ is constrained by the recent ATLAS
% result~\cite{atlas:chargedHiggs} for the top decay to charged Higgs
% and $b$-quark with the charged Higgs decay to $\tau$ and $\nu_{\tau}$.
%
Such a light charged Higgs boson is excluded by the recent ATLAS
search in the decays of top quarks~\cite{atlas:chargedHiggs}.
More generally, the charged Higgs boson mass smaller than 150 GeV and
$\tan\beta \lesssim 4$ has been excluded under the assumption of
Br$(H^+ \to \tau^+ \nu) = 1$.
%
% For such small $\tan\beta$, the charged Higgs boson dominantly decays into $\tau$
% and $\nu_{\tau}$,\footnote{
% For $\tan\beta=3$, the branching fractions of leading decay channels are given by
%   \begin{eqnarray}
%     \text{Br}(H^+\to\tau^+\nu_{\tau})&=&97\%,\nonumber\\
%     \text{Br}(H^+\to c\bar{s})&=&1.1\%,\nonumber\\
%     \text{BR}(H^+\to c\bar{b})&=&1.4\%,\nonumber
%   \end{eqnarray}
%   from the numerical calculation of \textsc{Nmssmtools}.
% }
% so the ATLAS result, which is shown in Fig. \ref{fig:atlas_chh1}, rules out such possibility.
% Setting aside the possibility of vanishing $c$, we would consider finite Higgs mixing case, i.e. larger $m_A$ and $m_{H^{\pm}}$ in the following paragraphs.
% **** more discussion ****
%
\begin{figure}[t!]
  \begin{center}
    \includegraphics[width=10cm]{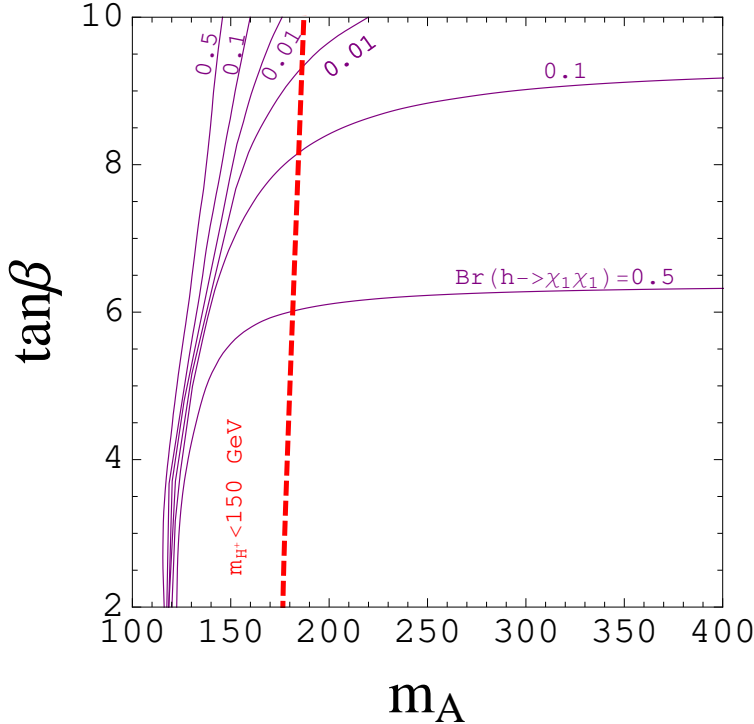}
    \caption{Contours of the branching fraction of the Higgs invisible
      decay for $\mu_{\text{eff}}$ and the GUT relation of gaugino
      masses. In the left side of red dashed region, charged Higgs
      mass is smaller than 150 GeV so that $\tan\beta\lesssim4$ is
      excluded by the charged Higgs search experiment.
      \label{fig:intertan}}
  \end{center}
\end{figure}
Fig.~\ref{fig:intertan} shows the result of our calculation
of the Higgs invisible decay ratio
depending on $m_A$ and $\tan\beta$ and the current LHC limit.
It should be mentioned that $m_A$ in the horizontal axis is the input
parameter at stop mass scale, so it is different from $m_A$ in
(\ref{eq:zeromix}) that is the value at the weak scale.
In addition, since the result includes loop corrections, it is
slightly different from what is expected from the tree-level
estimation. % in the previous paragraph.
In Fig.~\ref{fig:intertan}, one can find the region of $\tan\beta\gtrsim4$ and
$m_A\sim130$ GeV  in which the Higgs
invisible decay ratio becomes vanishingly small.
In this region, however, the production cross section and the decay branching
fraction of the lightest Higgs boson become different from the SM-like
one because of the CP-even Higgs mixing from the mass matrix (\ref{CPeven}).
In order to assess the viability of such non-SM-like Higgs scenario,
we would need more complicated study for whole SUSY parameters and
collider signatures. We will leave this for future work and focus on
the SM-like Higgs phenomenology from now on.

In the case of larger $m_A$, the invisible decay branching
fraction of the Higgs boson drastically increases so that larger
$\tan\beta$ is needed for the Higgs invisible decay to vanish as was
showed in (\ref{eq:g_Hchichi=0}).
For $m_A \gtrsim 200$ GeV, the invisible branching fraction becomes
nearly independent of $m_A$ since such region corresponds to the MSSM
decoupling limit, in which the lightest CP-even Higgs boson becomes
SM-like, so we find $c\simeq-\tan\alpha\simeq\cot\beta$.
In other words, the Higgs coupling to a neutralino pair
(\ref{coupl:Hchichi2}) depends only on $\tan\beta$.
In this region, we need $\tan\beta \gtrsim 9$ for
Br$(h\to\widetilde{\chi}_1^0\widetilde{\chi}_1^0)<0.1$.
It should be noted that this result is obtained with the GUT
relation of gaugino masses, % as commented in Fig.~\ref{fig:intertan}
and if one considers the TeV mirage relation, the corresponding
$\tan\beta$ value becomes larger. For such large $\tan\beta$,
the NMSSM feature of the sizable $\lambda$ contribution
to the tree-level Higgs boson mass is lost since
$\lambda^2v^2\sin^22\beta$ in (\ref{eq:Higgs_tree}) is already small
compared to the $Z$ boson contribution $m_Z^2\cos^22\beta$.
Thus, it is required to have large loop corrections from the stop sector
as in the MSSM.
\begin{figure}[t!]
  \begin{center}
    % \subfigure[$g_{h\chi_1\chi_1}$]{
    \subfigure[]{
      \includegraphics[height=7cm]{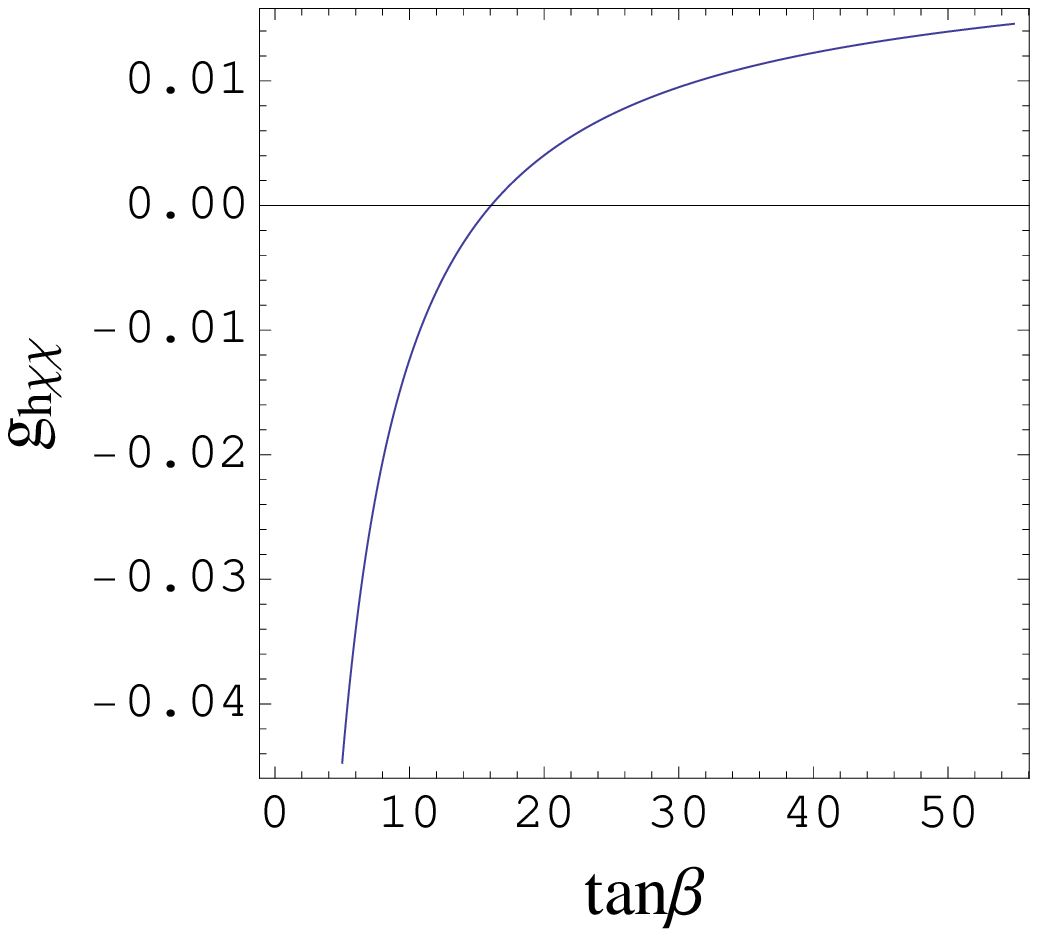}
    }
    % \subfigure[$\text{Br}(h\to\chi_1\chi_1)$]{
    \subfigure[]{
      \includegraphics[height=7cm]{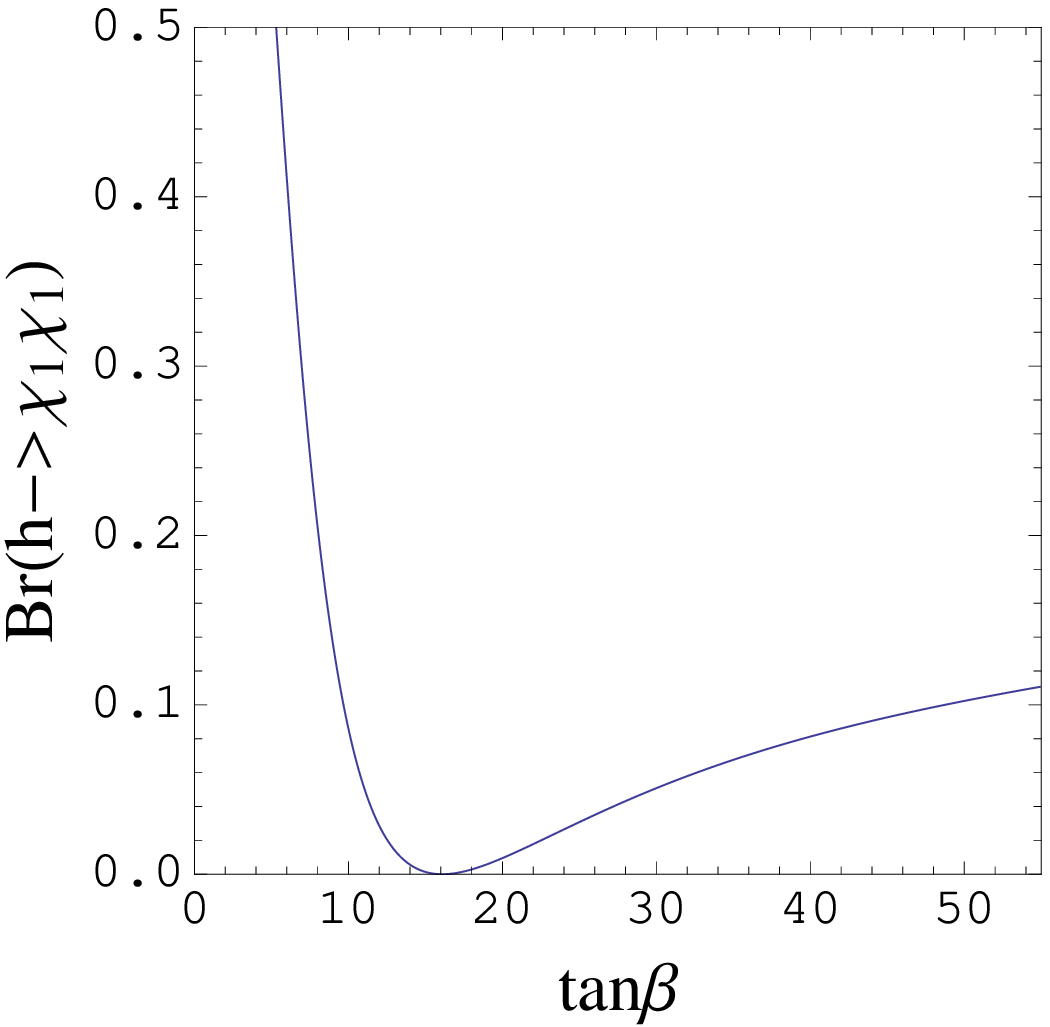}
    }
    \caption{(a) The lightest Higgs coupling to the lightest
      neutralino pair, (b) Branching fraction of the Higgs invisible
      decay.
      In these plots, we use $m_A=500$ GeV, $\mu_{\rm eff}=400$ GeV
      and the GUT relation of gaugino masses.
      \label{fig:Higgs_inv_gut}
    }
  \end{center}
\end{figure}
\begin{figure}[t!]
  \begin{center}
    % \subfigure[$g_{h\chi_1\chi_1}$]{
    \subfigure[]{
      \includegraphics[height=7cm]{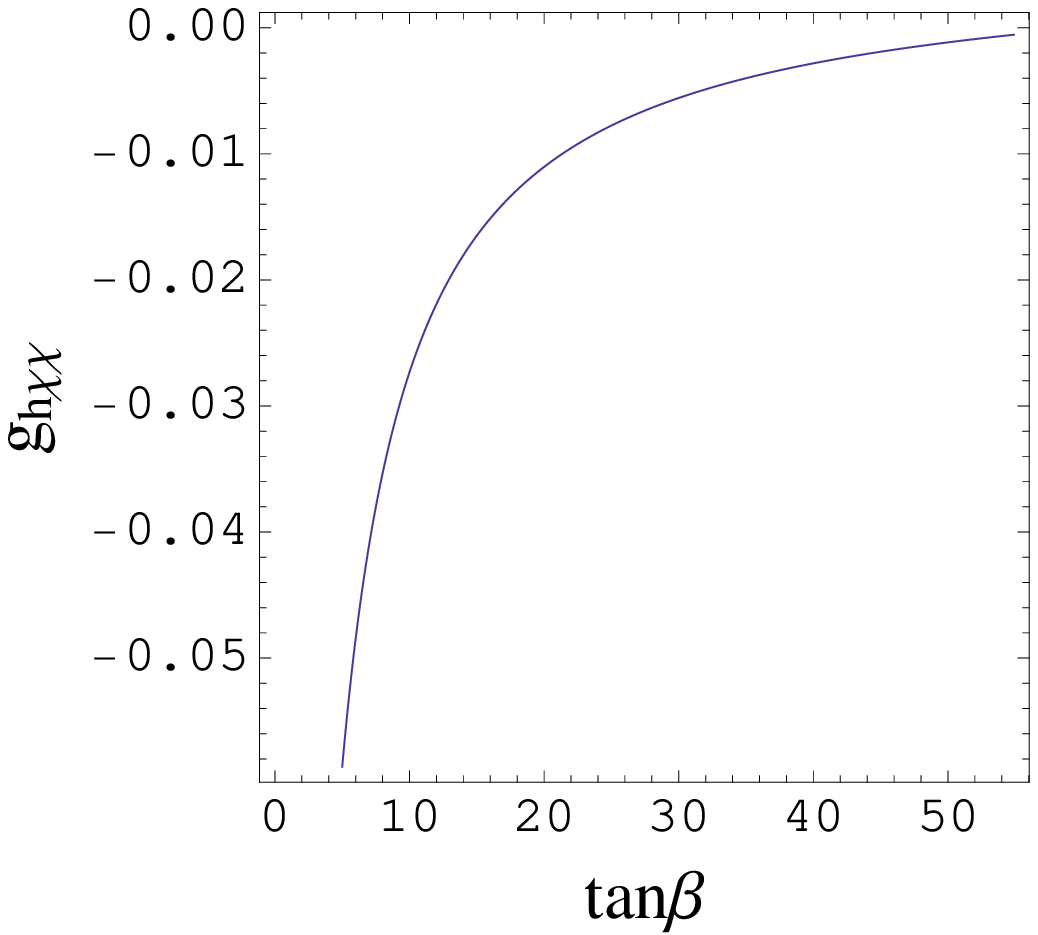}
    }
    % \subfigure[$\text{Br}(h\to\chi_1\chi_1)$]{
    \subfigure[]{
      \includegraphics[height=7cm]{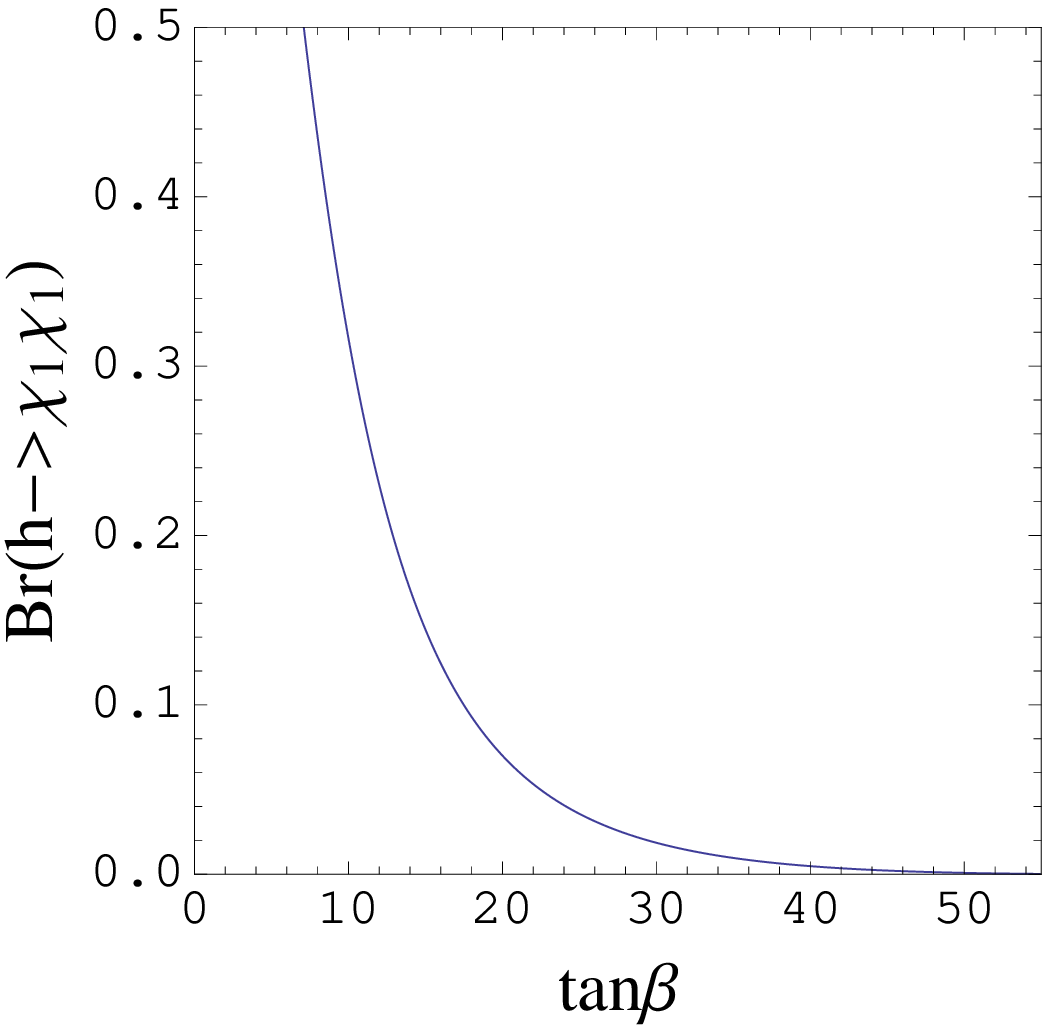}
    }
    \caption{(a) The lightest Higgs coupling to the lightest
      neutralino pair, (b) Branching fraction of the Higgs invisible
      decay.
      In these plots, we use $m_A=500$ GeV, $\mu_{\rm eff}=400$ GeV
      and the TeV mirage relation of gaugino masses.
      \label{fig:Higgs_inv_mirage2}
    }
  \end{center}
\end{figure}

For the comparison between cases of the GUT and TeV mirage scenarios,
we show the Higgs couplings to a neutralino pair and corresponding
branching fractions for both cases in Figs.~\ref{fig:Higgs_inv_gut}
and \ref{fig:Higgs_inv_mirage2}.
In the case of the GUT relation, the EW gauginos are
light ($M_1=160$ GeV and $M_2=330$ GeV), so the coupling vanishes
when $\tan\beta\sim12$ as can be seen in Fig.~\ref{fig:Higgs_inv_gut}(a).
This is consistent with % the fact that we discussed in
the relation (\ref{eq:g_Hchichi=0}).
Therefore, the branching fraction of the Higgs invisible decay is very
small near $\tan\beta\sim12$ as shown in Fig.~\ref{fig:Higgs_inv_gut}(b).
On the other hand, in the case of TeV mirage relation,
the gauginos are relatively heavy ($M_1=M_2=1$ TeV), thus
(\ref{eq:g_Hchichi=0}) can be satisfied only for very large values of
$\tan\beta$.
In Fig.~\ref{fig:Higgs_inv_mirage2}(a), we cannot see the solution of
(\ref{eq:g_Hchichi=0}) in the range of $1<\tan\beta<55$.
Instead, the size of the Higgs coupling to a neutralino pair becomes
smaller as $\tan\beta$ becomes larger.
Hence, nearly vanishing Higgs invisible decay can be attained when
$\tan\beta\gtrsim20$ as shown in Fig.~\ref{fig:Higgs_inv_mirage2}(b).

So far, we have seen that rather large $\tan\beta$ ($9 \lesssim
\tan\beta \lesssim 18$ for the GUT and $\tan\beta\gtrsim20$
for the TeV mirage relation) is needed to reduce the Higgs invisible
decay branching fraction below 10\% level.
Even if we allow the branching fraction up to 50\% level, one needs
$\tan\beta\gtrsim6$ for the GUT and $\tan\beta\gtrsim10$ for
the TeV mirage relation. This suppresses
the NMSSM contribution to the tree-level Higgs boson mass
and there is no difference from the MSSM in the
naturalness point of view. However, an important difference arises
due to the existence of the very light neutralino.
As one can see in (\ref{eq:chi1_mass}), the lightest neutralino mass
is generally smaller than 15 GeV for $\tan\beta\gtrsim5$ leading to
 non-trivial implications  in cosmology and collider
signatures. If such a light singlino-like neutralino is the LSP, its
annihilation cross section is very small, and thereby overproducing
the dark matter density.
To avoid this problem, we need a non-standard
cosmological history, which will be discussed in Sec.~\ref{sec:DM}.
On the other hand, the singlino-like LSP modifies
the decay topology of the supersymmetric particles.
This will be discussed in Sec.~\ref{sec:collider}.

\section{Dark Matter}
\label{sec:DM}

Let us now discuss the dark matter cosmology of the singlino-like LSP
for two regions of small and large $\tan\beta$
that survived the various constraints from particle phenomenology.
For each region, the standard LSP dark matter obtained by thermal
freeze-out has a difficulty in satisfying
the cosmological density and/or the direct detection bound. Instead, the non-standard cosmology
driven by the axion supermultiplet, the PQ sector, can provide a viable range of the
parameter space. In the following discussions, we assume
that R-parity is conserved and the lightest neutralino, mostly singlino, is the LSP
disregarding a possibility of lighter axino or gravitino being the LSP.

\subsection{\boldmath Small $\tan\beta$ region ($1 \lesssim\tan\beta \lesssim 2$)}

\begin{figure}[t!]
  \begin{center}
    \subfigure[$\Omega^{\rm TH}_{\widetilde\chi^0_1}h^2
    $]{\includegraphics[width=7cm]{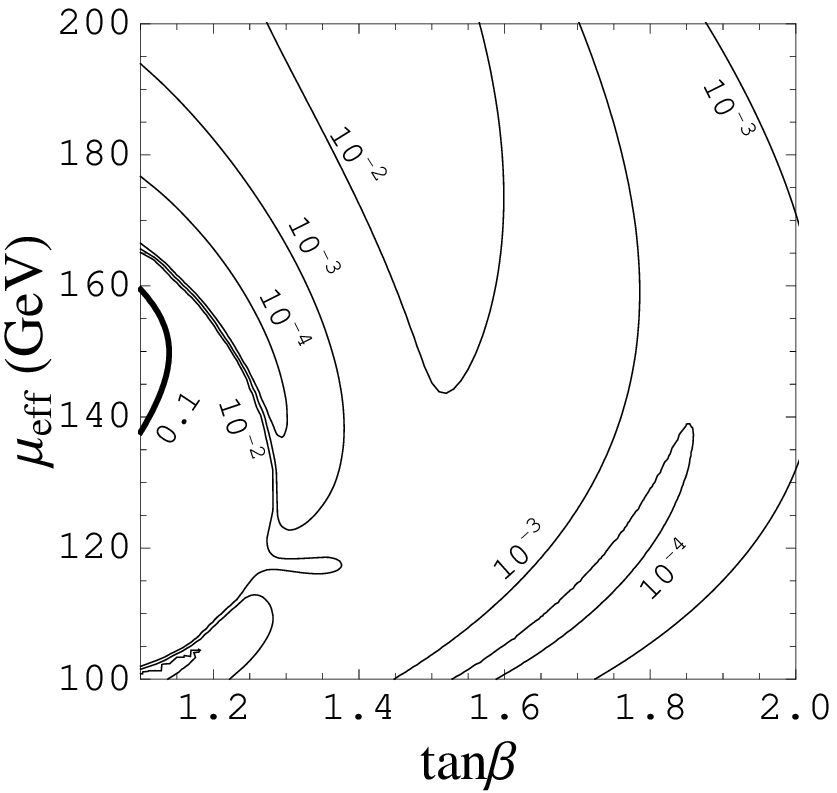}}
    \subfigure[$(\Omega_{\widetilde\chi^0_1}^{\rm TH}/\Omega_{\rm
      DM})\,\sigma_{\rm
      SI,\widetilde\chi^0_1}$]{\includegraphics[width=7cm]{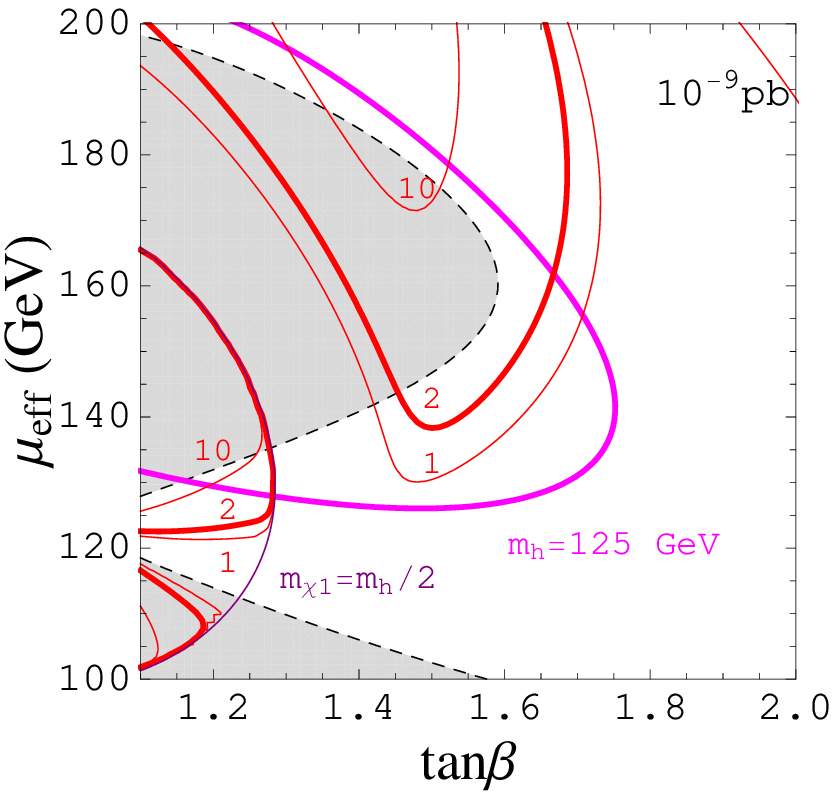}}
    \caption{Plot for low $\tan\beta$ with $6M_1=3M_2=M_3=1$ TeV.
      Other parameters are the same as those of Fig. \ref{fig:lowtan_gut}. (a) Black curves
      denote the thermal relic density of the neutralino LSP,
      $\Omega^{\rm TH}_{\widetilde\chi^0_1}h^2$. (b)
      Red curves denote the central values of the nucleonic scattering cross-section of the lightest neutralino, $(\Omega_{\widetilde\chi^0_1}^{\rm TH}/\Omega_{\rm DM})\sigma_{\rm SI,\widetilde\chi^0_1}$ in the unit  of $10^{-9}$ pb.
      The magenta curve denotes 125 GeV Higgs mass,
      the purple curve denotes
      $m_{\widetilde\chi^0_1}=m_{h}/2$, and the gray-shaded region is excluded by OPAL.
      For $50\ {\rm GeV} \lesssim m_{\widetilde\chi^0_1} \lesssim 70\ {\rm GeV}$, the XENON100 rules out the region
      $(\Omega_{\widetilde\chi^0_1}^{\rm TH}/\Omega_{\rm DM})\,\sigma_{\rm SI,\widetilde\chi^0_1} >2\times 10^{-9}$ pb at 90\% C.L.~\cite{XENON:2012nq}.}
    \label{fig:lowtan_relic_gut}
  \end{center}
\end{figure}
\begin{figure}[t!]
  \begin{center}
    \subfigure[$\Omega^{\rm TH}_{\widetilde\chi^0_1}h^2
    $]{\includegraphics[width=7cm]{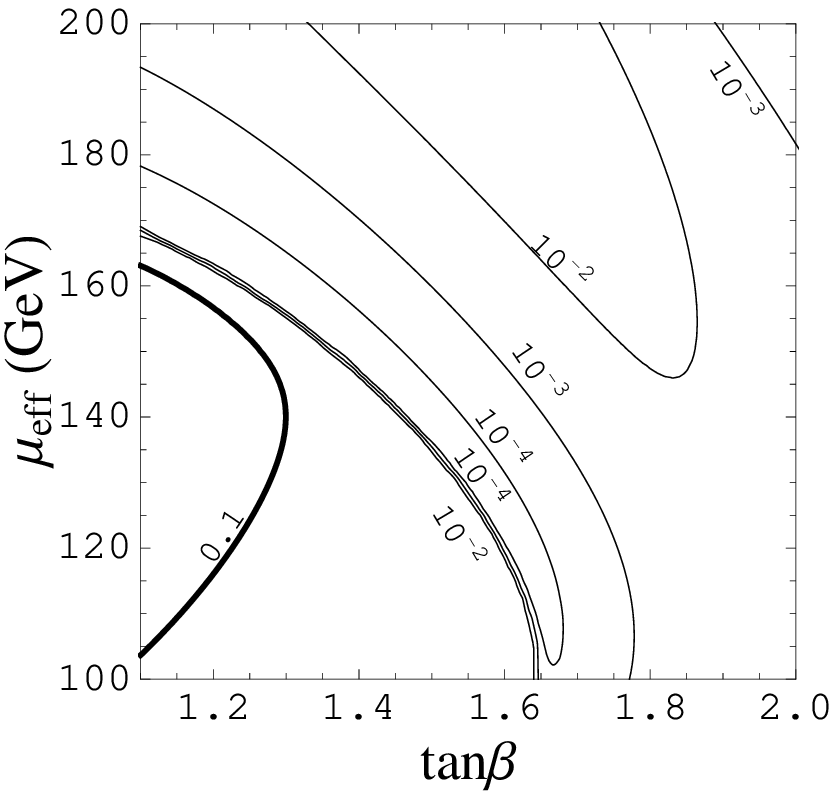}}
    \subfigure[$(\Omega_{\widetilde\chi^0_1}^{\rm TH}/\Omega_{\rm
      DM})\,\sigma_{\rm SI,
      \widetilde\chi^0_1}$]{\includegraphics[width=7cm]{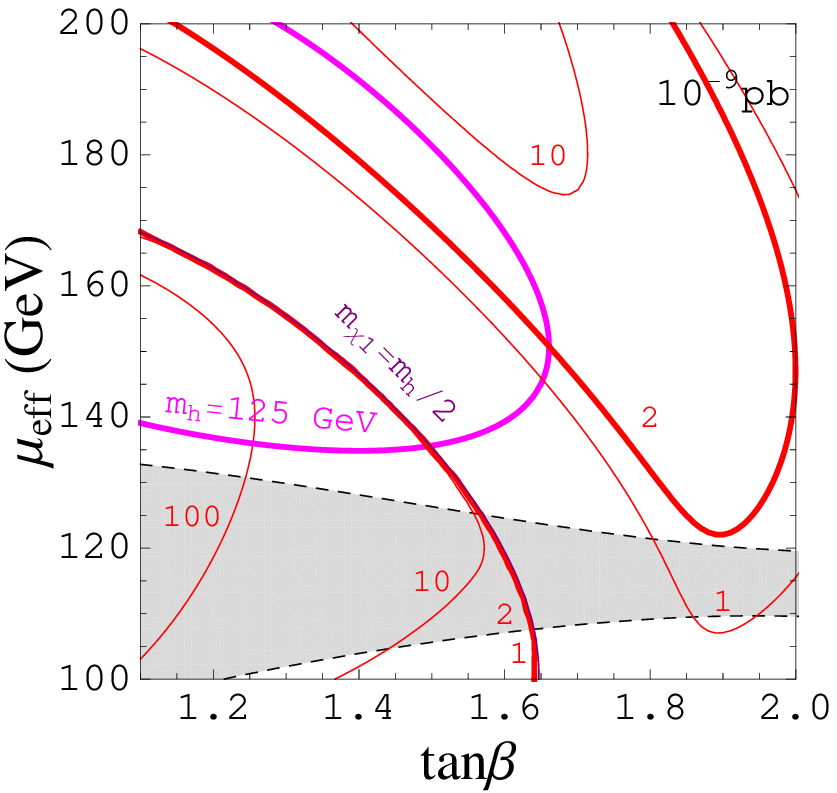}}
    \caption{Plot for low $\tan\beta$ with $M_1=M_2=M_3=1$ TeV.
      Other parameters are the same as those of Fig. \ref{fig:lowtan_gut}. (a) Black curves
      denote the relic amount of the neutralino LSP, $\Omega^{\rm TH}_{\widetilde\chi^0_1}h^2$. (b)
      Red curves denote the central values of the nucleonic scattering cross-section of the lightest neutralino, $(\Omega_{\widetilde\chi^0_1}^{\rm TH}/\Omega_{\rm DM})\,\sigma_{\rm SI,\widetilde\chi^0_1}$ in the unit  of $10^{-9}$ pb.
      The magenta curve denotes 125 GeV Higgs mass,
      the purple curve denotes
      $m_{\widetilde\chi^0_1}=m_{h}/2$, and the gray-shaded region is excluded by OPAL.
      For $ 50\  {\rm GeV}\lesssim m_{\widetilde\chi^0_1} \lesssim 70\ {\rm GeV}$, the XENON100 rules out the region
      $(\Omega_{\widetilde\chi^0_1}^{\rm TH}/\Omega_{\rm DM})\sigma_{\rm SI,\widetilde\chi^0_1} >2\times 10^{-9}$ pb at 90\% C.L.~\cite{XENON:2012nq}.
      \label{fig:lowtan_relic_mirage}
    }
  \end{center}
\end{figure}

In the small $\tan\beta$ region, the LSP mass $m_{\widetilde\chi^0_1}$
is around $50 - 70$ GeV.
As studied in the context of the nMSSM~\cite{Menon:2004wv},
 both
$s$-channel $Z$ boson exchange ($\widetilde\chi^0_1\widetilde\chi^0_1
\rightarrow Z \rightarrow f \bar f$) and $s$-channel Higgs boson
exchange ($\widetilde\chi^0_1\widetilde\chi^0_1\rightarrow h
\rightarrow f\bar f$) are equally important for the neutralino
annihilation cross-section.
The effect of the Higgs exchange is more important than in the case of the MSSM
since the $h$-$\widetilde\chi^0_1$-$\widetilde\chi^0_1$ coupling
(\ref{ghxx}) is enhanced compared to that of the MSSM.
The resulting relic abundance $\Omega^{\rm TH}_{\widetilde\chi^0_1}h^2$
is represented in the left panels of Figs.~\ref{fig:lowtan_relic_gut} and
\ref{fig:lowtan_relic_mirage} for  the
GUT scale unified gaugino masses and the TeV scale mirage mediation, respectively.
The superscript ``TH" implies the dark matter density produced from thermal freeze-out.
In each figure, the
narrow contour lines of  $\Omega^{\rm TH}_{\widetilde\chi^0_1}h^2$
around $10^{-2}-10^{-4}$ for smaller $\tan\beta$ correspond to the
Higgs resonance region, $m_{\widetilde\chi^0_1}\approx m_{h}/2$.
As $\tan\beta$ increases on the right-hand side of the Higgs
resonance, $m_{\widetilde\chi^0_1}< m_{h}/2$,
$m_{\widetilde\chi^0_1}$ approaches $m_{Z}/2$ and thus
 $\Omega^{\rm TH}_{\widetilde\chi^0_1}h^2$ decreases again due to the $Z$ resonance.
As $\tan\beta$ decreases on the left-hand side of the Higgs resonance,
$m_{\widetilde\chi^0_1}> m_{h}/2$, the neutralino relic abundance
increases and can reach the correct amount of the dark matter,
$\Omega^{\rm TH}_{\widetilde\chi^0_1}h^2\approx 0.11$.

Right panels of Figs.~\ref{fig:lowtan_relic_gut} and
\ref{fig:lowtan_relic_mirage} represent the effective
spin-independent elastic scattering cross-section  between
the nucleon and the dark matter in units of $10^{-9}\ {\rm pb}$.
For the direct detection bound on the scattering cross-section \cite{XENON:2012nq},
the standard local dark matter density
is assumed, that is, $n_{\rm X}
=\rho^{\rm loc}_{\rm DM}/m_{\widetilde\chi^0_1}
\simeq 0.3\  {\rm cm}^{-3}( {\rm GeV}/m_{\widetilde\chi^0_1})$ corresponding to
$\Omega_{\rm DM} h^2 \approx 0.11$.
However, in our parameter region where the thermal relic abundance
of the neutralino dark matter  is smaller than $\Omega_{\rm DM}h^2\approx 0.11$,
the neutralino number density around the Earth
deceases by the factor of $\Omega^{\rm TH}_{\widetilde\chi^0_1}/\Omega_{{\rm DM}}$.
Thus, we show the effective cross-section between the nucleon
and the assumed dark matter, $(\Omega^{\rm
  TH}_{\widetilde\chi^0_1}/\Omega_{{\rm DM}}) \times \sigma_{\rm
  SI,\widetilde\chi^0_1}$, where the spin-independent elastic
scattering cross-section between the nucleon and the neutralino
$\sigma_{\rm SI,\widetilde\chi^0_1}$ is given by
\bea
\sigma_{\rm SI,\widetilde\chi^0_1} =\frac{4}{\pi}\left(\frac{m_{\widetilde\chi^0_1} m_N}
  {m_N + m_{\widetilde\chi^0_1}}\right)^2 f_N^2 .
\eea
Here, $N$ represents the nucleon $(p,\,n)$ and the constant $f_N$ is
\bea \frac{f_N}{m_N}= \sum_{q=u,d,s}
f_{Tq}^{(N)}\left(\frac{\alpha_q}{m_q}\right)+
\frac{2}{27}\sum_{q=c,b,t}f^{(N)}_{Tg}\left(\frac{\alpha_q}{m_q}\right),
\eea where $m_q$ denotes the quark mass,
$\alpha_q$ is the coupling constant of the quark level effective Lagrangian term
${\cal L}_{\rm eff}=
\alpha_q {\bar {\widetilde\chi}^0_1}\widetilde\chi^0_1 \bar q q$, and
 the hadronic matrix elements are given by $m_Nf^{(N)}_{Tq}=\langle
N|m_q\bar q q|N\rangle$ and
$f^{(N)}_{Tg}=1-\sum_{q=u,d,s}f^{(N)}_{Tq}$~\cite{Shifman:1978zn}.
The numerical values of the matrix element $f^{N}_{Tq}$ are
determined in \cite{Ellis:2000ds} \bea \label{formfactor}
&&f^{(p)}_{Tu}=0.020\pm 0.004,\quad f^{(p)}_{Td}=0.026\pm
0.005,\quad f^{(p)}_{Ts}=
0.118\pm 0.062, \nonumber\\
&&f^{(n)}_{Tu}=0.014\pm 0.003,\quad f^{(n)}_{Td}=0.036\pm 0.008,\quad f^{(n)}_{Ts}=
0.118\pm 0.062. \eea
In our case,
 the coupling constant  $\alpha_q$  is
dominantly given by $t$-channel Higgs exchange diagram,
\bea
\alpha_q &\simeq & -\frac{ g_{h\widetilde\chi^0_1\widetilde\chi^0_1}}{m_{h}^2}
\left(\frac{g_2 m_q S_{11}}{m_W\sin\beta}\right) \eea
for the up-type quarks. For the down-type quark, $\alpha_q$ is
obtained by the appropriate replacements ($S_{11}\rightarrow S_{21}$,
$\sin\beta\rightarrow \cos\beta$).
The Higgs-LSP-LSP coupling $g_{h\widetilde\chi^0_1\widetilde\chi^0_1}$ is given by
\bea\label{ghxx}
&&g_{h\widetilde\chi^0_1\widetilde\chi^0_1}=
-g_2 \Big(N_{12}-\tan\theta_W N_{11}\Big)
\Big(S_{11} N_{13}- S_{12} N_{14}\Big)\nonumber\\
&&\hskip1.8cm+\ \sqrt{2}\lambda \Big(S_{13}N_{13}N_{14}
+ N_{15}(S_{12}N_{13}+S_{11}N_{14})\Big).
\eea
The first line in (\ref{ghxx}) is the same as in the MSSM, and
the second line comes from the superpotential term, $\lambda S H_u H_d $.
Due to the sizable Higgsino component of the LSP,
the second term dominates $g_{h\widetilde\chi^0_1\widetilde\chi^0_1}$.
The recent XENON100 data puts an upper limit on the effective
spin-independent LSP-nucleon cross-section around $2\times 10^{-9}\
{\rm pb}$ for the mass range of $50 -70$ GeV at 90\% C.L.~\cite{XENON:2012nq}.
Note that the region that gives $\Omega_{\widetilde\chi^0_1}^{\rm TH}h^2=0.11$
is far above the XENON100 bound.
However, for the points on the 125 GeV
Higgs line close to the Higgs resonance point, the XENON100 bound can
be avoided because the neutralino thermal relic density can be much
below the observed dark matter density.
More specifically, we find that the mass difference between $m_{\widetilde\chi^0_1}$ and $m_{h}/2$
should be smaller than  ${\cal O}(0.1)$ GeV,
\bea \label{narrowband}
m_{\widetilde\chi_1^0} - \frac{m_{h}}{2} \lesssim 0.1\ {\rm GeV}. \eea
Recall that, since the neutralino contribution to the dark matter density is small,
the major component of the dark matter
density can come from the axion for $v_{\rm PQ} \sim 10^{12}$ GeV in our scenario,
\bea
\Omega_a h^2 \simeq  0.23 \left(\frac{v_{\rm PQ}}{10^{12}\ {\rm GeV}}\right)^{7/6}
\langle \theta^2\rangle,
\eea
where $\theta$ is the initial misalignment angle, typically of order one.

Let us note that the PQ-NMSSM has a late-time decaying saxion field $s$,
a CP-even partner of the axion $a$, which can dilute away the LSP relic density
calculated previously.
If the PQ sector
is stabilized by SUSY breaking effects as discussed in Sec.~\ref{sec:PQ}, the saxion typically gets a
mass of the order of the weak scale,
and its interaction with other particles are suppressed by
$1/v_{\rm PQ}$, so it has a long lifetime. It is natural
to have a period in the early universe during which
the energy density of the universe is dominated by the coherent oscillation of $s$ or
vacuum energy of the PQ sector field. At the end of this period
($t\sim 1/\Gamma_s$ where
$\Gamma_s $ is the total decay rate of $s$),
the saxion will decay to produce radiation reheating the Universe at the temperature $T_{\rm RH}$
obtained from the relation $\rho_{s}(T_{\rm RH})\simeq \rho_{\rm r}(T_{\rm RH})$.
The reheat temperature $T_{\rm RH}$ is given by
 \bea \label{TRH}
T_{\rm RH}\simeq 200\ {\rm MeV} \left(\frac{10}{g_\ast(T_f)}\right)^{1/4}
\left(\frac{0.1}{{\rm Br}_{a}}\right)^{1/2}
\left(\frac{m_s}{100\ {\rm GeV}}\right)^{3/2}
\left(\frac{10^{12}\ {\rm GeV}}{v_{\rm PQ}}\right),
\eea
where ${\rm Br}_{a}$ is the branching fraction
of the saxion decay into the axion pair having the rate
$\Gamma_{s\rightarrow a a}=m_{s}^3/(64\pi v_{\rm PQ}^2)=
\Gamma_s {\rm Br}_a$~\cite{Choi:1996vz}. For $v_{\rm PQ} = 10^{12}$ GeV,
the reheat temperature is much smaller than the freeze-out
temperature of the LSP  ($T_{\rm RH} \ll T_f\simeq
m_{\widetilde\chi^0_1}/22\sim 3$ GeV),
and thus the depleted thermal LSP population cannot be regenerated.
Furthermore, the non-thermal production of the LSP
from the saxion decay can also be forbidden if
the saxion mass $m_s$ is taken to be smaller than $2m_{\widetilde\chi^0_1}$.
Therefore, the stringent constraint (\ref{narrowband}) on the LSP mass can be relaxed.

\subsection{\boldmath Large $\tan\beta$ region ($5\lesssim\tan\beta $)}

In the large $\tan\beta$ region, we found that
the lightest neutralino mass is below ${\cal O}(5\ {\rm GeV})$
from the consideration of the Higgs invisible decay.
Since $m_{\widetilde\chi^0_1}$ is far from the Higgs resonance region,
the LSPs are annihilated into the SM fermion pairs
dominantly by the $s$-channel $Z$ boson exchange,
$\widetilde\chi^0_1\widetilde\chi^0_1\rightarrow Z\rightarrow f\bar f$.
In the mass range of $1\ {\rm GeV} \lesssim
m_{\widetilde\chi^0_1}\lesssim 5$ GeV,
the LSP relic abundance is given by
\bea\label{TH1}
\Omega^{\rm TH}_{\widetilde \chi^0_1}h^2 &=& 1.07\times 10^9
\left(\frac{m_{\widetilde \chi^0_1}}{\sqrt{8\pi}M_{\rm Pl}}\right)
\left( \int^{T_f}_0 dT  g_\ast(T)^{1/2} \langle\sigma_{\rm ann}
  v\rangle_{T}\ {\rm GeV}^2\right)^{-1}\\
&= &  10^3\gamma_{f\widetilde\chi^0_1}\left(\frac{x_f}{9}\right)^2
\left(\frac{10}{g_\ast(T_f)} \right)^{1/2}
\left(1-\frac{4m_{\widetilde \chi^0_1}^2}{m_Z^2}\right)^2
\left(\frac{2\ {\rm GeV}}{m_{\widetilde \chi^0_1}}\right)^2
\left(\frac{0.04}{|N_{13}^2-N_{14}^2|}\right)^2, \nonumber
\eea where $T_f$ is the freeze-out temperature of $\widetilde\chi^0_1$,
$x_f=m_{\widetilde \chi^0_1}/T_f$ is around $8\sim 10$ ($15$) for
$m_{\widetilde\chi^0_1}\sim 1\ {\rm GeV}$ ($5\ {\rm GeV}$),
and $g_\ast(T_f)\simeq 10$ for $T_f\sim 100\ {\rm MeV}$. The order-one constant
$\gamma_{f\widetilde\chi^0_1}$ basically counts the number of fermions lighter than
the LSP. The neutralino mixing elements (\ref{N_13}) and (\ref{N_14}) give
\bea
|N_{13}^2 - N_{14}^2| \simeq \frac{\lambda^2 v^2
  \sin^2\beta}{\mu_{\rm eff}^2}
\simeq0.04\left(\frac{m_{\widetilde\chi^0_1}}{2\ {\rm GeV}}\right)
\left(\frac{\tan\beta}{20}\right)\left(\frac{500\ {\rm GeV}}{\mu_{\rm eff}}\right)
\eea
for large $\tan\beta$. Recall that $|N_{13}^2- N_{14}^2|$ is
bounded by 0.13 as in (\ref{Zinvbound}).
Therefore, in order to avoid the overclosure dark matter density,
the neutralino abundance has to be depleted
by the factor of $\Delta \sim 10^4-10^2$ for
$m_{\widetilde\chi^0_1}\sim 1-5$ GeV.

In the large $\tan\beta$ region, $m_{\widetilde\chi^0_1}$ is sensitive
to $\lambda$ while the Higgs mass is not, and can be much smaller than
1 GeV if we allow for $\lambda$ smaller than the nominal choice 0.7.
For instance, one finds
$m_{\widetilde\chi^0_1} \sim 10\ {\rm MeV}$ for $\lambda \sim 0.1$.
In this case, the freeze out temperature $T_f$ is larger than $m_{\widetilde\chi^0_1}$ and
becomes a few ${\rm GeV}$. The relic abundance is then given by
\bea
\Omega^{\rm TH}_{\widetilde\chi^0_1}h^2 \simeq 0.3\times 10^5 \left(\frac{m_{\widetilde\chi^0_1}}
  {30\ {\rm MeV}}\right)\left(\frac{100}{g_\ast(T_f)}\right).
\eea
Thus, even larger dilution factor $\Delta\sim 10^5$ is needed for $m_{\widetilde\chi^0_1}={\cal O}(10\ {\rm MeV})$.

It is amusing to note that the dilution mechanism by the saxion field
discussed in the previous subsection can successfully deplete the LSP thermal abundance
as well as produce the right amount of the non-thermal LSP relic density.
Since the singlino LSP is light enough to be produced by the saxion decay, we need to check
how sizable amount of the LSP dark matter can be produced in this process.
Such a non-thermal production is controlled by the ratio
$m_{\widetilde{\chi}_1^0}/m_s$ as the
$s$-$\widetilde{\chi}_1^0$-$\widetilde{\chi}_1^0$ coupling is
proportional to $m_{\widetilde{\chi}_1^0}/v_{\rm PQ}$ whereas the
$s$-$a$-$a$ coupling is proportional to $m_s/v_{\rm PQ}$.
We here consider the case $m_s\sim 100\ {\rm GeV}$,
then the dominant saxion decay mode can be $s\rightarrow b\bar b$ whose coupling is proportional to
$c_{sH} m_b \tan\beta/ v_{PQ}$, where $c_{sH}$ is an order-one
parameter representing the saxion-Higgs mixing.
For $\tan\beta\gtrsim 25$, the $s$-$b$-$\bar b$ coupling can be
larger than the $s$-$a$-$a$ coupling allowing for Br$_a < 0.1$.
Now, one can find the  relic abundance of non-thermally
produced LSP through the decay $s\rightarrow \widetilde\chi^0_1\widetilde\chi^0_1$ as follows.
\bea\label{NTHofN1}
\Omega^{\rm NTH}_{\widetilde\chi^0_1 }h^2
&=& 0.1  \gamma_{s\widetilde\chi^0_1 \widetilde\chi^0_1}
\left(\frac{m_{\widetilde\chi^0_1}}{30\,{\rm MeV}}\right)^3
\left(\frac{100\,{\rm GeV}}{m_s}\right)
\left(\frac{T_{\rm RH}}{10\, {\rm MeV}}\right),
\eea
where $\gamma_{s\widetilde\chi^0_1 \widetilde\chi^0_1}$ is
the order-one constant controlling the $s$-$\widetilde\chi^0_1$-$\widetilde\chi^0_1$ coupling.
The above relation shows that the light singlino LSP abundance can be
in the right range for $m_{\widetilde\chi^0_1} \sim 30\ {\rm MeV}$
(obtainable for, e.g., $\lambda\simeq 0.07$, $\tan\beta =25$, $\mu_{\rm eff}=300\ {\rm GeV}$),
$m_s\sim 100$ GeV, and $v_{\rm PQ}\sim 10^{13}$ GeV.
Since we used the PQ symmetry breaking scale $v_{\rm PQ}$ larger than
$10^{12}$ GeV, we also have to consider the axion dark matter abundance.
The entropy dumping from the saxion decay at
$T_{\rm RH}={\cal O}(10)$ MeV can dilute the axion
relic density as well. In this case, the axion dark matter at low
decay temperature is given by~\cite{Choi:1996vz}
\bea
\Omega_a h^2 \simeq 3\times 10^{-3}\left(\sqrt{\frac{g_{\ast}(T_{\rm RH})}{10}}
  \left(\frac{T_{\rm RH}}{10\ {\rm MeV}}\right)^2\right)^{0.98}
\left(\frac{v_{\rm PQ}}{10^{13}\ {\rm GeV}}\right)^{1.5}
\left(\frac{200\ {\rm MeV}}{\Lambda_{\rm QCD}}\right)^{1.9}
\langle \theta^2\rangle,
\eea
which is negligible.

\section{Collider signature at the LHC}
\label{sec:collider}

In this section, we study the collider signature of the PQ-NMSSM
taking some benchmark parameter points 
in the case of the TeV scale mirage relation of gaugino masses
discussed in the previous sections.
The mass spectra and branching ratios of the sparticles and Higgs
bosons have been calculated with the modified codes of
\textsc{Nmssmtools}, and are given in Tables~\ref{tab:massA}
and \ref{tab:decayA} in the small $\tan\beta$ scenario,
and in Tables~\ref{tab:massB}  and \ref{tab:decayB} in the large $\tan\beta$ scenario,
respectively.
Notice that the stop mass is chosen to be as light as  $\sim 500$ GeV
consistently with the 125 GeV Higgs boson for the small $\tan\beta$ benchmark point,
whereas it has to be heavy ($\sim 1$ TeV) with a large mixing for the large $\tan\beta$
benchmark point.

\begin{table}[tb!]
  \caption{Sparticle and Higgs masses (in GeV) for 
  the small $\tan\beta$ benchmark point.}
  \label{tab:massA}
  \begin{center}
    \begin{tabular}{c c c c c c c c c c c c c}
      \hline\hline&&&&&&\\[-2mm]
      $\widetilde g$ & $\widetilde u_L$ & $\widetilde u_R$ & $\widetilde t_1$ &
      $\widetilde t_2$ & $\widetilde b_1$ & $\widetilde b_2$ &
      $\widetilde e_L$ & $\widetilde e_R$ & $\widetilde\tau_1$ & $\widetilde\tau_2$
      &&
      \\[2mm]
      \hline&&&&&&&&\\[-2mm]
      1119 & 2033 & 2033 & 492 &
      542 & 504 & 506 &
      2000 & 2000 & 2000 & 2000
      &&
      \\[2mm]
      \hline\hline&&&&&&&&&&&\\[-2mm]
      $\widetilde\chi_1^0$ & $\widetilde\chi_2^0$ & $\widetilde\chi_3^0$  &
      $\widetilde\chi_4^0$ & $\widetilde\chi_5^0$ & $\widetilde\chi_1^\pm$ &
      $\widetilde\chi_2^\pm$ & $H_1$ & $H_2$ & $H_3$ & $A_1$ & $A_2$ & $H^\pm$
      \\[2mm]
      \hline&&&&&&&&&&\\[-2mm]
      63  & 134  & 206 &
      992 & 1030  & 129 &
      1029 & 125  & 347 &  609 & 339 & 618 & 341
      \\[2mm]
      \hline\hline
    \end{tabular}
  \end{center}
  \smallskip
  \caption{Main decay modes for the sparticles and the lightest
    Higgs boson and their branching fractions in \% for the small
    $\tan\beta$ benchmark point.}
  \label{tab:decayA}
  \begin{center}
    \begin{tabular}{lr | lr | lr}
      \hline\hline&&&&&\\[-2mm]
      $\widetilde g \to \widetilde t_2 t$       & 32.0 &
      $\widetilde t_2 \to \widetilde\chi_1^\pm b$ & 32.8 &
      $\widetilde\chi_1^\pm \to \widetilde\chi_1^0 W^\ast$ & 100.0
      \\[2mm]
      $\widetilde g \to \widetilde t_1 t$      & 16.7 &
      $\widetilde t_2 \to \widetilde\chi_2^0 t$ & 26.1 &
      $\widetilde\chi_2^0 \to \widetilde\chi_1^0 Z^\ast (H^\ast)$ & 99.8
      \\[2mm]
      $\widetilde g \to \widetilde b_2 b$      & 25.8 &
      $\widetilde t_2 \to \widetilde\chi_1^0 t$ & 27.6 &
      $H_1 \to b \bar{b}$ & 62.9
      \\[2mm]
      $\widetilde g \to \widetilde b_1 b$      & 25.5 &
      $\widetilde t_1 \to \widetilde\chi_1^\pm b$ & 46.9 &
      $H_1 \to WW^\ast$ & 19.2
      \\[2mm]
      &&
      $\widetilde t_1 \to \widetilde\chi_2^0 t$ & 14.9 &
      $H_1 \to ZZ^\ast$ & 2.1
      \\[2mm]
      &&
      $\widetilde t_1 \to \widetilde\chi_1^0 t$ & 19.0 &
      $H_1 \to \gamma\gamma$ & 0.2
      \\[2mm]
      &&
      $\widetilde b_2 \to \widetilde\chi_1^\pm t$ & 99.8 &
      &
      \\[2mm]
      &&
      $\widetilde b_1 \to \widetilde\chi_1^\pm t$ & 99.1 &
      &
      \\[2mm]
      \hline\hline
    \end{tabular}
  \end{center}
\end{table}

\begin{table}[tb!]
  \caption{Sparticle and Higgs masses (in GeV) for the large $\tan\beta$ benchmark point.}
  \label{tab:massB}
  \begin{center}
    \begin{tabular}{c c c c c c c c c c c c c}
      \hline\hline&&&&&&\\[-2mm]
      $\widetilde g$ & $\widetilde u_L$ & $\widetilde u_R$ & $\widetilde t_1$ &
      $\widetilde t_2$ & $\widetilde b_1$ & $\widetilde b_2$ &
      $\widetilde e_L$ & $\widetilde e_R$ & $\widetilde\tau_1$ & $\widetilde\tau_2$
      &&
      \\[2mm]
      \hline&&&&&&&&\\[-2mm]
      1166 & 2031 & 2031 & 837 &
      1158 & 1014 & 1036 &
      1001 & 1001 &  995 & 1007
      &&
      \\[2mm]
      \hline\hline&&&&&&&&&&&\\[-2mm]
      $\widetilde\chi_1^0$ & $\widetilde\chi_2^0$ & $\widetilde\chi_3^0$  &
      $\widetilde\chi_4^0$ & $\widetilde\chi_5^0$ & $\widetilde\chi_1^\pm$ &
      $\widetilde\chi_2^\pm$ & $H_1$ & $H_2$ & $H_3$ & $A_1$ & $A_2$ & $H^\pm$
      \\[2mm]
      \hline&&&&&&&&&&\\[-2mm]
      0.03 &  298 & 307 &
        987 & 1023 & 301 &
       1023 &  123 & 683 & 2510 & 683 & 2510 & 687
      \\[2mm]
      \hline\hline
    \end{tabular}
  \end{center}
  \smallskip
  \caption{Main decay modes for the sparticles and the lightest
    Higgs boson and their branching fractions in \% for the large
    $\tan\beta$ benchmark point.}
  \label{tab:decayB}
  \begin{center}
    \begin{tabular}{lr | lr | lr}
      \hline\hline&&&&&\\[-2mm]
      $\widetilde g \to \widetilde t_1 t$         & 75.5 &
      $\widetilde t_1 \to \widetilde\chi_1^\pm b$ & 37.8 &
      $\widetilde\chi_1^\pm \to \widetilde\chi_1^0 W$ & 100.0
      \\[2mm]
      $\widetilde g \to \widetilde b_2 b$      & 11.0 &
      $\widetilde t_1 \to \widetilde\chi_2^0 t$ & 35.5 &
      $\widetilde\chi_2^0 \to \widetilde\chi_1^0 Z$ & 56.0
      \\[2mm]
      $\widetilde g \to \widetilde b_1 b$ & 13.5 &
      $\widetilde t_1 \to \widetilde\chi_1^0 t$ & 0.1 &
      $\widetilde\chi_2^0 \to \widetilde\chi_1^0 H_1$ & 44.0
      \\[2mm]
      &&
      $\widetilde b_2 \to \widetilde t_1 W$ & 17.2 &
      $H_1 \to b\bar{b}$ & 67.1
      \\[2mm]
      &&
      $\widetilde b_2 \to \widetilde\chi_1^\pm t$ & 55.6 &
      $H_1 \to WW^\ast$ & 15.9
      \\[2mm]
      &&
      $\widetilde b_2 \to \widetilde\chi_2^0 b$ & 12.2 &
      $H_1 \to ZZ^\ast$ & 1.6
      \\[2mm]
      &&
      $\widetilde b_2 \to \widetilde\chi_1^0 b$ & $\simeq$ 0.0 &
      $H_1 \to \gamma\gamma$ & 0.2
      \\[2mm]
      &&
      $\widetilde b_1 \to \widetilde t_1 W$ & 29.0 &
      &
      \\[2mm]
      &&
      $\widetilde b_1 \to \widetilde\chi_1^\pm t$ & 60.4 &
      &
      \\[2mm]
      &&
      $\widetilde b_1 \to \widetilde\chi_2^0 b$ & 6.0 &
      &
      \\[2mm]
      &&
      $\widetilde b_1 \to \widetilde\chi_1^0 b$ & $\simeq$ 0.0 &
      &
      \\[2mm]
      \hline\hline
    \end{tabular}
  \end{center}
\end{table}

In the small $\tan\beta$ scenario, the main decay topologies are
similar to those of the typical MSSM scenario with light stops and sbottoms
due to the large mixing between the singlino and the Higgsino.
On the other hand, in the large $\tan\beta$ scenario, the existence of
a very light singlino LSP prevents the direct decay of the
sparticles into the LSP and thus there appear additional decay steps
compared to the similar MSSM decay processes.
The main decay processes of the large $\tan\beta$ benchmark point are
\begin{itemize}
  \item $\widetilde g \to \widetilde t_1 t \to \widetilde\chi_1^\pm b t
    \to \widetilde\chi_1^0 W b t$,
  \item $\widetilde g \to \widetilde t_1 t \to \widetilde\chi_2^0 t t
    \to \widetilde\chi_1^0 Z\,(H_1)\, t t$,
  \item $\widetilde g \to \widetilde b_1 b \to \widetilde\chi_1^\pm t b
    \to \widetilde\chi_1^0 W t b$,
  \item $\widetilde g \to \widetilde b_1 b \to \widetilde\chi_2^0 b b
    \to \widetilde\chi_1^0 Z\,(H_1)\, b b$.
\end{itemize}
Hereafter the SM-like Higgs boson $h$ is denoted by $H_1$
representing the lightest one among three CP-even Higgs bosons.
As one can see from the above decay chains,
the final states consist dominantly of multi-jet coming from
the multi-top/bottom, and $H_1$/$Z$/$W$ $+$ missing energy that
may escape the early LHC searches on the channels with (two to four)
jets and large missing energy~\cite{Khachatryan:2011tk}.\footnote{
This was also claimed in \cite{Das:2012rr,Vasquez:2012hn} in the
context of $\mathbb{Z}_3$-invariant NMSSM. Here, we stress that
the relatively light singlino-like LSP can be naturally attained
in the PQ-NMSSM. } Furthermore, the
heavier neutralinos will eventually decay into the light singlino
LSP, which makes the missing energy generically small as claimed
recently in \cite{Lisanti:2011tm}. This will reduce the efficiency
of the searches on the channels with large jet multiplicities and
missing energy~\cite{Aad:2011qa, Aad:2012hm}.
Another notable feature is that the relatively light
stops and sbottoms lead to top/bottom-rich signal events.
The recent constraints on the stop and sbottom masses were set by the
searches on the gluino-mediated production~\cite{ATLAS:2012pq} and the
direct production of stops and sbottoms~\cite{Aad:2011cw,
  cite:stop}. In our study, the gluino and squarks are
chosen be heavy enough to evade such search results.

In order to study the collider phenomenology,
 we have generated Monte Carlo (MC)
event samples for a proton-proton collision at 14 TeV by
\textsc{Herwig++ 2.5.2}~\cite{Bahr:2008pv}.
The generated event samples have been scaled to 10 fb$^{-1}$ integrated
luminosity.
For the sake of the simple analysis, we consider only the leading-order
cross sections calculated with \textsc{Herwig++} and the
CTEQ6L~\cite{Pumplin:2002vw} parton distribution functions (PDF).
The generator-level event samples have been further processed with the fast
detector simulation program \textsc{Delphes 2.0}~\cite{Ovyn:2009tx}
using the ATLAS detector card.
Jets are reconstructed using the anti-$k_t$ jet clustering
algorithm~\cite{Cacciari:2008gp} with radius parameter of 0.4.
Isolated electrons (muons) are required to have the transverse
momentum $p_{\rm T}>20$ (10) GeV and the pseudo-rapidity $|\eta|<2.47$
(2.4).
In the recent version of \textsc{Delphes}, the missing transverse
momenta $\sla{\bf p}_{\rm T}$ are defined as the negative vector sum of the
transverse momenta of all the calorimetric cells and muon candidates.
On top of that, we resolve overlaps between jets with $|\eta|<2.8$
and leptons by following the recent ATLAS analysis~\cite{Aad:2011qa}.
Jets lying within a distance $\Delta R \equiv \sqrt{(\Delta\eta)^2 +
  (\Delta\phi)^2} <0.2$ from an electron are discarded.
Then, any lepton remaining within a distance $\Delta R < 0.4$ from such a
jet is discarded. Finally, all jets with $|\eta|>2.8$ are discarded.
From now on, we will use only the remaining electrons, muons, and jets
for the analysis.

The dominant SM backgrounds for
the signal with large jet multiplicities come from the top-pair process
and $W$ or $Z$ bosons produced in association with jets. The di-boson and the
single-top processes can contribute to the backgrounds as well, but
they are negligible. For a simple study of the
background effects, we here consider only the top-pair process.
Since our study mainly concerns the signal events with a large number
of jets, we generate fully hadronic, semi-leptonic, and fully leptonic $t\bar{t}$
events with up to two additional partons in the matrix
element using \textsc{Alpgen
  2.14}~\cite{Mangano:2002ea} and CTEQ6L PDF sets.
The parton showering to generate additional jets, and the
fragmentation and hadronization are performed by \textsc{Herwig++}.
The MC samples have been processed with the \textsc{Delphes} to
reconstruct jets and isolated leptons and adjust the detector effects.

To suppress the backgrounds, we impose the basic event selection cuts as follows.
\begin{enumerate}[(i)]
  \item At least 6 jets with $p_{\rm T} > 80$ GeV,
  \item no isolated electron or muon,
  \item missing transverse energy $\sla{E}_{\rm T} > 80$ GeV,
  \item ${\cal S}_{\rm T} > 0.2$.
\end{enumerate}
Here, ${\cal S}_{\rm T}$ is the transverse sphericity defined as
\begin{eqnarray}
  {\cal S}_{\rm T} \equiv \frac{2\lambda_2}{\lambda_1 + \lambda_2} ,
\end{eqnarray}
where $\lambda_1$ and $\lambda_2$ are the eigenvalues of the
$2\times 2$ sphericity tensor ${\cal S}_i^j=\sum_{k} p_{ki} p^{kj}$ of
the reconstructed objects~\cite{Aad:2009wy}. This variable is
known to be useful to suppress QCD events in which back-to-back
configurations (${\cal S}_{\rm T}\sim 0$) are dominated. Although
such backgrounds are not considered here, this variable will be
included in the analysis.
In the ATLAS MC study for the inclusive SUSY search~\cite{Aad:2009wy},
the azimuthal angular separation cut, $\Delta\phi({\rm jet} - \sla{\bf
  p}_{\rm T})$, is imposed further to reduce the jet
mis-measurement effect. However, it has been recently noted that this cut
variable may lose the efficiency if the next-to-lightest neutralino
$\widetilde\chi_2^0$ was boosted and the lightest Higgs boson $H_1$ or $Z$
boson decayed into $b\bar{b}$ aligned with the $\sla{\bf p}_{\rm T}$
vector~\cite{Vasquez:2012hn}.
We also note that the cut on the missing transverse energy is
required to reduce the fully hadronic $t\bar{t}$ background process,
even though it should be relatively milder than the typical SUSY searches
because of the small missing energy.

\begin{figure}[t!]
  \begin{center}
    \subfigure[]{
      \includegraphics[width=7cm]{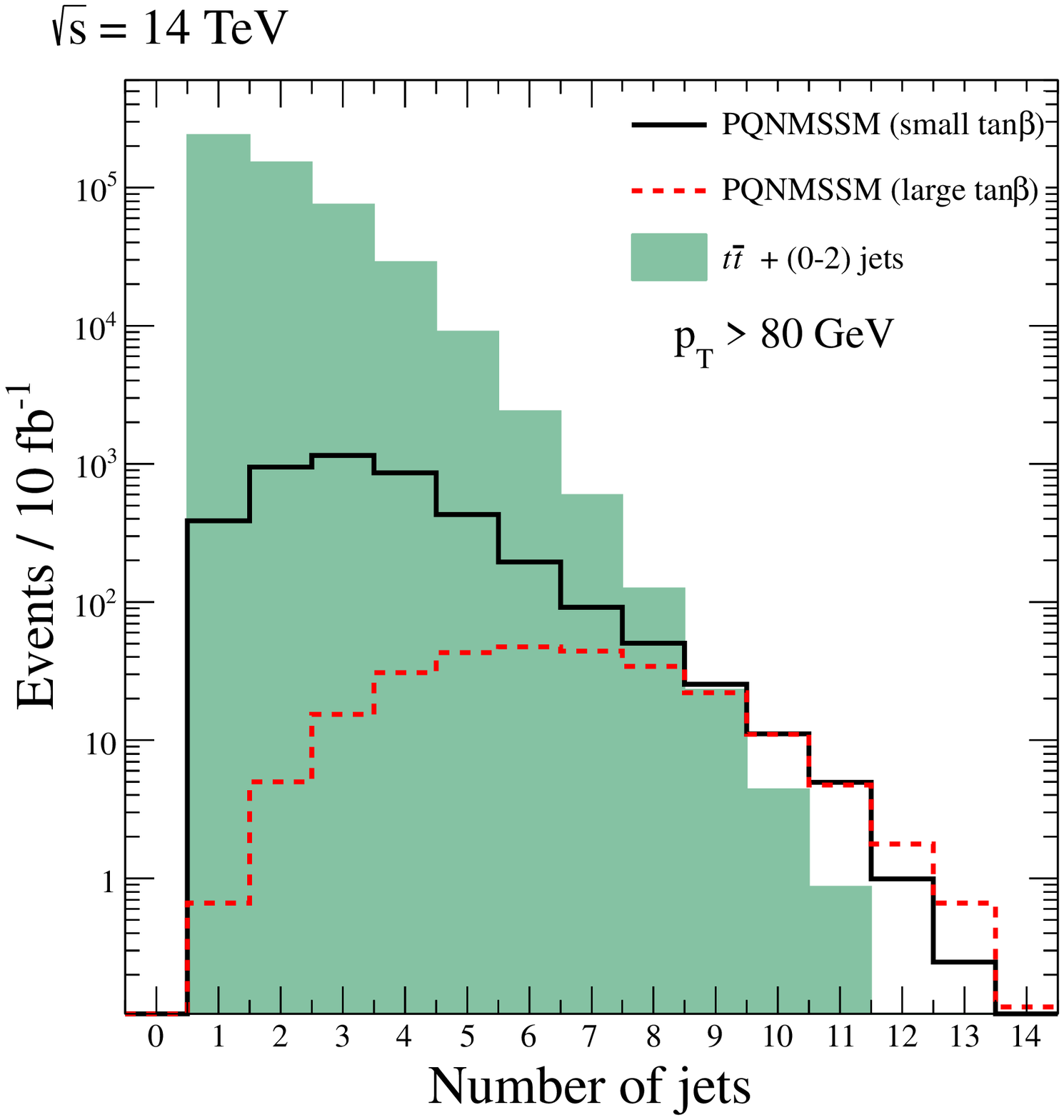}
    }
    \subfigure[]{
      \includegraphics[width=7cm]{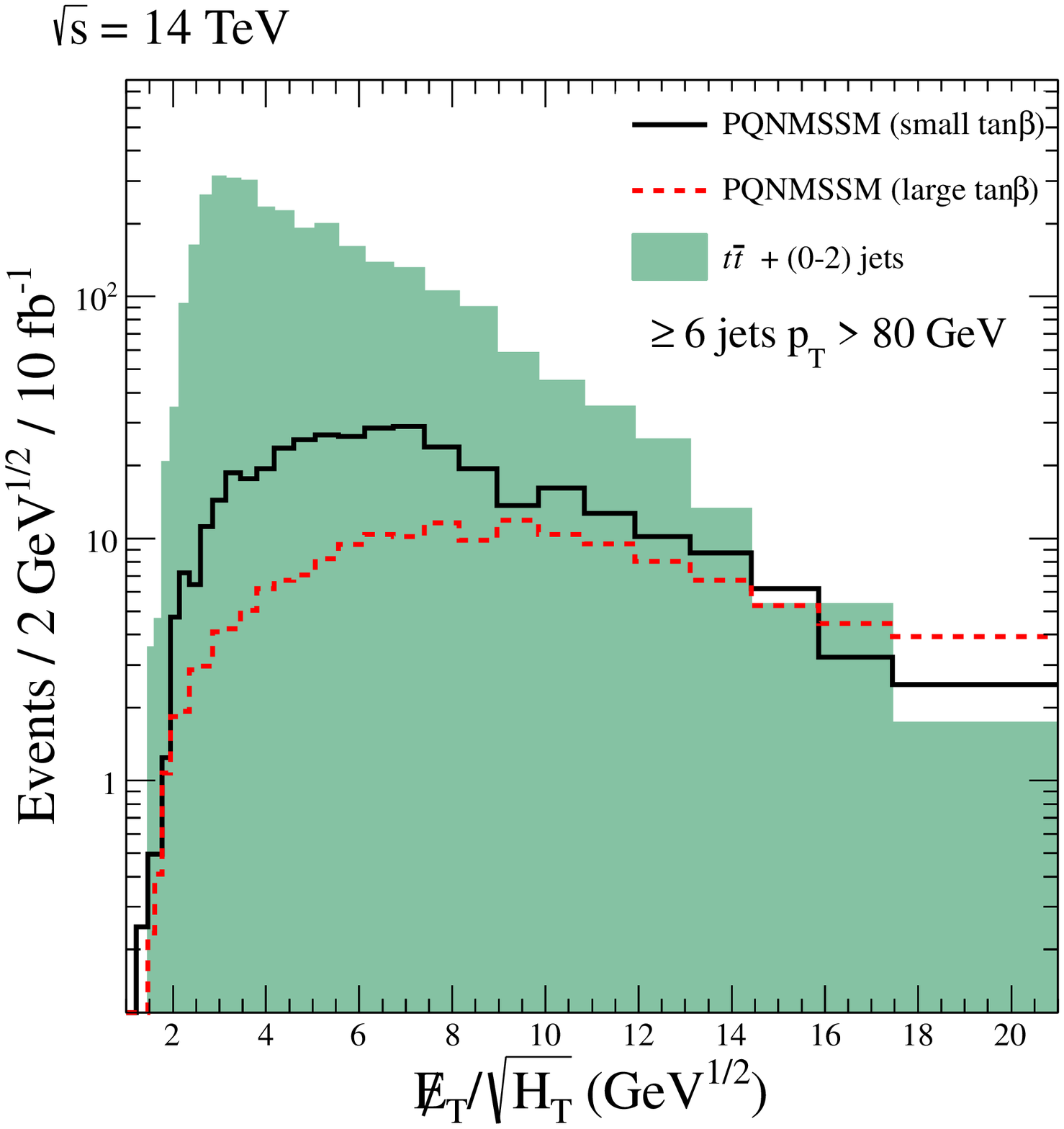}
    }
  \end{center}
  \caption{The distributions of (a) the jet multiplicities
    and (b) the $\sla{E}_{\rm T} / \sqrt{H_{\rm T}}$ for the signal
    events and $t\bar{t} +$ jets.}
  \label{fig:NumJet1_MetHT}
\end{figure}

We show the jet multiplicity distributions for the signals and
backgrounds after applying the basic selection cuts in
Fig.~\ref{fig:NumJet1_MetHT}(a).
 One can see that there are relatively more number of energetic
  jets in the large $\tan\beta$ scenario. 
  This is because the additional decay step caused by the existence of
  the singlino LSP as well as the fact that the sparticle production
  modes are mostly consisted by a gluino-pair production as the squarks
  are too heavy to be produced. On the other hand, the $\widetilde t
  \widetilde t$ and $\widetilde b \widetilde b$ production cross
  sections are somehow sizable in the small $\tan\beta$ scenario. 
  The relatively short decay steps result in the smaller number of jets.
% It is mainly due to $\widetilde t\widetilde t$ or $\widetilde b \widetilde b$ productions in
% the small $\tan\beta$ scenario and the additional decay step caused by the
% existence of the singlino LSP as discussed above.}
%
Fig.~\ref{fig:NumJet1_MetHT}(b) shows the
distribution of the ratio $\sla{E}_{\rm T} / \sqrt{H_{\rm T}}$, where
$H_{\rm T}$ is the scalar sum of the transverse momenta of all jets
with $p_{\rm T} > 50$ GeV and $|\eta| < 2.8$.
This variable has been known to be useful to increase the performance
of the missing energy reconstruction~\cite{Aad:2012hm}. We require the
condition of $\sla{E}_{\rm T} / \sqrt{H_{\rm T}}$ being larger than 4
GeV$^{1/2}$ in addition to the basic event selection cuts.
\begin{figure}[t!]
  \begin{center}
    \subfigure[]{
      \includegraphics[width=7cm]{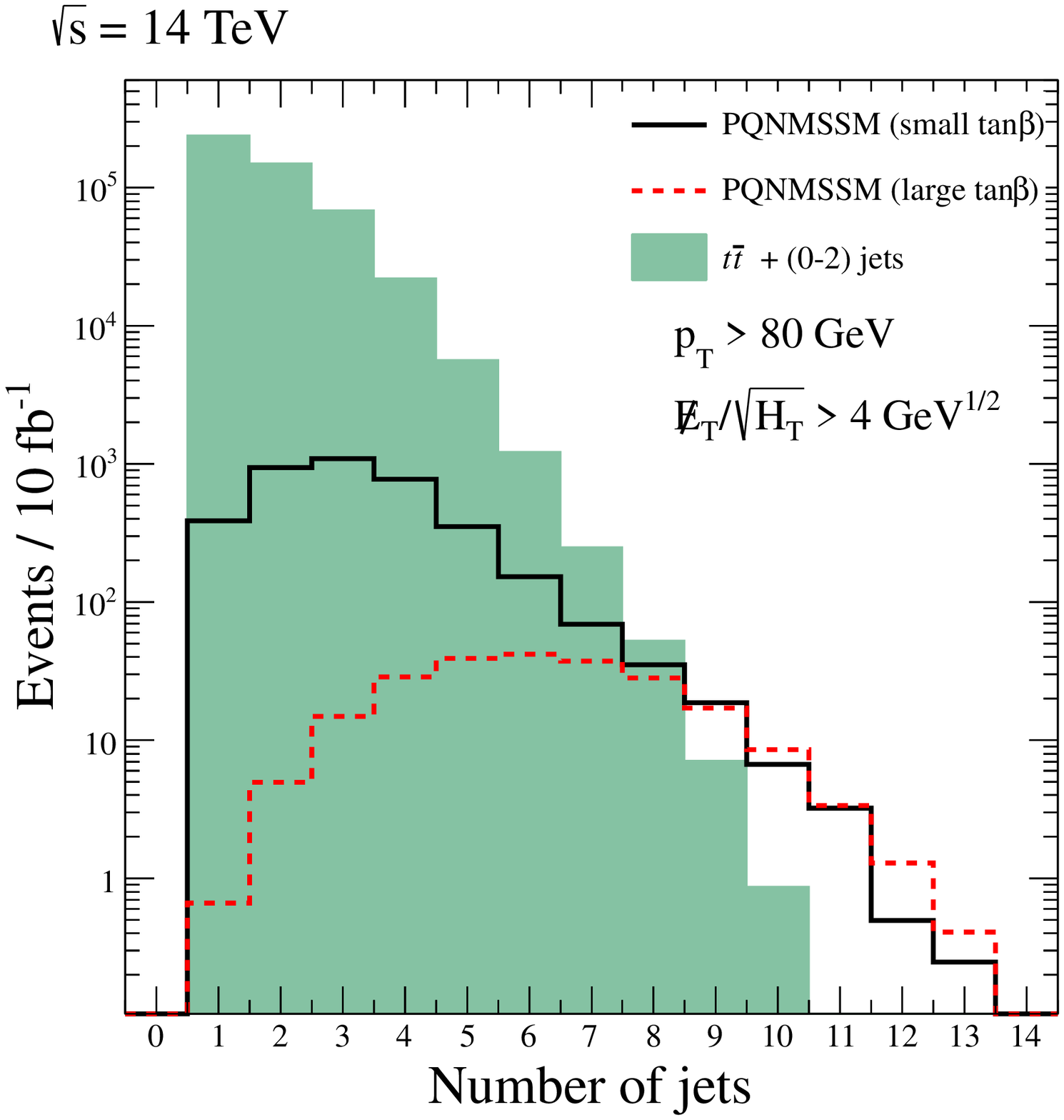}
    }
    \subfigure[]{
      \includegraphics[width=7cm]{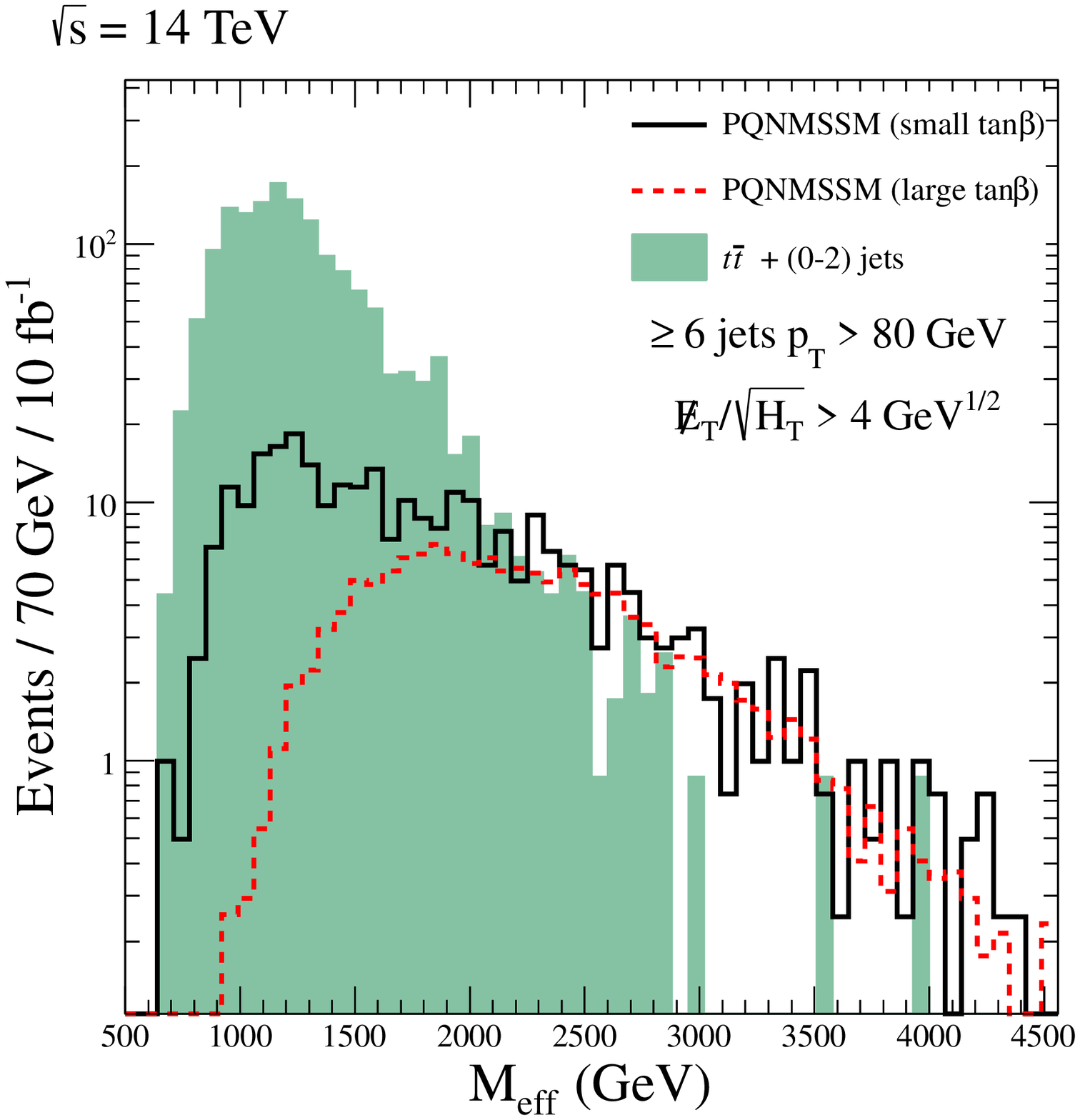}
    }
  \end{center}
  \caption{The distributions of (a) the jet multiplicities
    and (b) the $M_{\rm eff}$ for the signal
    events and $t\bar{t} +$ jets with imposing the $\sla{E}_{\rm T} /
    \sqrt{H_{\rm T}}$ cut.}
  \label{fig:NumJet2_Meff}
\end{figure}
By employing this cut variable on top of the basic selections cuts,
the jet multiplicity distributions are shown in
Fig.~\ref{fig:NumJet2_Meff}(a). One can see that the background has
been reduced, while the signals remain almost untouched in the
region of the large jet multiplicity.
In Fig.~\ref{fig:NumJet2_Meff}(b), we also show the
distribution of the effective mass defined as
\begin{eqnarray}
  M_{\rm eff} \equiv \sla{E}_{\rm T} +
  \sum_{{\rm jets}} p_{\rm T},
\end{eqnarray}
where the summation is over all jets with $p_{\rm T} > 50$ GeV and
$|\eta| < 2.8$ in the event. The $M_{\rm eff}$ has been known to be a
good variable for discriminating SUSY signal events from SM
backgrounds~\cite{Tovey:2002jc}. In the literature, it was noted that
the peak position of the $M_{\rm eff}$ distribution has a strong
correlation with the SUSY mass scale. In the case of the small
$\tan\beta$ point, the $M_{\rm eff}$ distribution is peaked in the
lower position than in the case of the large $\tan\beta$
point. This is because of a sizable amount of direct production
rates of the light stops and sbottoms, whereas the signal events for the large
$\tan\beta$ point come practically from the gluino-pair process.

For a crude estimation of the signal significance, we show how the
cross sections of the signals and the backgrounds change under
each event selection cut in Table~\ref{table:cut}.
\begin{table}[tb!]
  \caption{Cut flows of the signals and backgrounds in fb.}
  \label{table:cut}
  \begin{center}
    \begin{tabular}{l | cc ccc}
      \hline\hline&&&&&\\[-2mm]
      \multirow{2}*{Selection cuts} &
      \multicolumn{2}{c}{PQ-NMSSM} &
      \multirow{2}*{$t\bar{t}$} &
      \multirow{2}*{$t\bar{t} + 1$ jet} &
      \multirow{2}*{$t\bar{t} + 2$ jets}
      \\[2mm]
      & small $\tan\beta$ & large $\tan\beta$ &
      &&
      \\[2mm]
      \hline&&&&&\\[-2mm]
      $\geq 6$ jets with $p_{\rm T} > 80$ GeV &
      105.7 & 57.3 &
      450.2 & 1650.6  &  2055.8
      \\[2mm]
      Lepton veto &
      \,\,51.8 & 20.5 &
      359.0 & 1259.8 & 1528.9
      \\[2mm]
      $\sla{E}_{\rm T} > 80$ GeV &
      \,\,44.7 & 19.4 &
      \,\,29.7 & \,\,148.1 & \,\,219.1
      \\[2mm]
      ${\cal S}_{\rm T} > 0.2$ &
      \,\,38.0 & 16.6 &
      \,\,23.9 & \,\,119.6 & \,\,173.0
      \\[2mm]
      $\sla{E}_{\rm T}/\sqrt{H_{\rm T}} > 4$ GeV$^{1/2}$ &
      \,\,28.6 & 13.8 &
      \,\,12.7 & \,\,\,\,58.8 & \,\,\,\,82.8
      \\[2mm]
      $M_{\rm eff} > 1000$ GeV &
      \,\,26.2 & 13.8 &
      \,\,\,\,8.4 & \,\,\,\,45.9 & \,\,\,\,67.4
      \\[2mm]
      $M_{\rm eff} > 1500$ GeV &
      \,\,16.6 & 12.4 &
      \,\,\,\,1.0 & \,\,\,\,11.5 & \,\,\,\,19.5
      \\[2mm]
      \hline\hline
    \end{tabular}
  \end{center}
\end{table}
Although we did not perform a more accurate estimation which requires
higher-order cross sections, optimization of the
cut values by the multivariate analysis techniques, and 
understanding the systematic uncertainties, 
we can expect the $5\sigma$ discovery for both
PQ-NMSSM scenarios with several fb$^{-1}$ of the
integrated luminosity.

\section{Conclusions}
\label{sec:concl}

Motivated by the axion solution of the strong CP problem and the recent discovery of a 125 GeV
Higgs boson at the LHC, we investigated phenomenological consequences of the PQ symmetry
realized in the context of NMSSM.  The minimal form of the PQ-NMSSM, in which
the singlino mass comes only from the singlino-Higgsino mixing and
thus the LSP from the singlino-Higgsino sector becomes lighter than
about 70 GeV, is shown to be tightly constrained
by the Higgs invisible decay if the LSP is a candidate of the dark
matter and it relic density is determined by the standard freeze-out
process.

Taking $\lambda=0.7$, which remains perturbative up to the PQ scale for $\tan\beta$ close to 1,
the 125 GeV Higgs can be obtained for stop mass around 500 GeV.
Such a small $\tan\beta$ is favored as the Higgs invisible decay can be forbidden kinematically
($m_{\widetilde{\chi}_1^0}> m_h/2$).
However, it requires a large Higgsino component for the LSP and thus
the recent XENON100 bound on the nucleonic cross-section of the LSP at 90\% C.L.
excludes almost all the parameter region except a narrow band of the LSP mass close to the Higgs
resonance point ($m_{\widetilde{\chi}_1^0} - m_h/2<0.1$ GeV) which suppresses the
thermal LSP relic density significantly.  We also note that a late decay of the saxion inherent
in the model can produce a huge amount of entropy diluting away the thermal LSP abundance.
Then, the severe constraint on the LSP mass is invalidated.
For larger $\tan\beta$, the LSP mass becomes smaller opening the Higgs invisible decay.
In this case, a cancellation can be arranged to suppress the Higgs-LSP-LSP coupling.
Combined with the charged Higgs mass bound at the LHC,
 the Higgs invisible decay branching fraction is shown to become smaller
 than 10\% for $\tan\beta > 9$.  In this parameter region, the NMSSM contribution
to the Higgs boson mass is negligible and thus a large stop mass $\sim 1$ TeV is needed to get the
Higgs mass of 125 GeV. As the LSP is almost purely singlino for large
$\tan\beta$, its thermal freeze-out density is orders of magnitude
larger than required.
However, it turns out that a late decay of the saxion, washing out the
dangerous thermal LSP relics again, can produce the right amount of
non-thermal dark matter population for the singlino mass of order 10
MeV with $\lambda\sim 0.1$.

Generic collider signatures of the PQ-NMSSM with a light singlino-like
LSP are multi-jets from multi-top/bottom and $h/Z/W$ plus missing
energy in the final states. Taking two benchmark points for small and
large $\tan\beta$, we analyzed such signals at the 14 TeV LHC to find
that a 5$\sigma$ discovery is possible for a few fb$^{-1}$ of the
integrated luminosity.

We finally note that  the unsatisfactory dark matter properties applied to the minimal
PQ-NMSSM in the standard cosmology can be evaded in a more general PQ-NMSSM allowing a
suitable bare singlino mass term.

%%%%%%%%%%%%%%%%%%%%%%%

\vspace{-0.2cm}
\subsection*{Acknowledgements}
\vspace{-0.3cm}
KJB, SHI, CBP, and CSS gratefully acknowledge the hospitality of KIAS
where part of this work was carried out. KJB is supported by TJ Park
Postdoctoral Fellowship of POSCO TJ Park Foundation. KC and SHI are
supported by the National Research Foundation of Korea(NRF) grant
funded by the Korea government(MEST)
(No. 2007-0093865 and No. 2012R1A2A2A05003214). KJB, KC, and SHI are also
supported by the BK21 project by the Korean Government.
CBP is partially supported by the grants FPA2010-17747,
Consolider-CPAN (CSD2007-00042) from the MICINN,
HEPHACOS-S2009/ESP1473
from the C. A. de Madrid and the contract ``UNILHC''
PITN-GA-2009-237920 of the European Commission.
CSS is supported by Basic Science Research Program through the
National Research Foundation of Korea (NRF) funded by the Ministry of
Education, Science and Technology (No. 2011-0011083).
CSS acknowledges the Max Planck Society (MPG), the Korea Ministry of
Education, Science and Technology (MEST), Gyeongsangbuk-Do and Pohang
City for the support of the Independent Junior Research Group at the Asia Pacific
Center for Theoretical Physics (APCTP).

%%%%%%%%%%%%%%%%%%%%%%%

\section*{Appendix}
\appendix

\section{Neutralino mixing matrices} \label{sec:Neutmixing}
The neutralino mass matrix is given in (\ref{Neutmatrix}),
% \begin{equation} \nonumber
%   {\cal M}_{\widetilde{\chi}^0}=
%   \begin{pmatrix}
%     M_1 & 0 & -g_1v_d/\sqrt{2} & g_1v_u/\sqrt{2} & 0 \\
%     & M_2 & g_2v_d/\sqrt{2} & -g_2v_u/\sqrt{2} & 0 \\
%     & & 0 & -\mu_{\text{eff}} & -\lambda v_u \\
%     & & & 0 & -\lambda v_d \\
%     & & & & 0
%   \end{pmatrix}
% \end{equation}
% The above matrix
which can be diagonalized by the method of perturbative
diagonalization as in \cite{Bae:2007pa},
\begin{equation}
  {\cal M}^{\text{diag}}=N{\cal M}_{\widetilde{\chi}^0}N^T=VU{\cal
    M}_{\widetilde{\chi}^0}U^TV^T=V{\cal M}V^T,
\end{equation}
where
\begin{equation}
  U=
  \begin{pmatrix}
    1 & 0 & 0 & 0 & 0\\
    0 & 1 & 0 & 0 & 0 \\
    0 & 0 & \frac{1}{\sqrt{2}} & -\frac{1}{\sqrt{2}} & 0 \\
    0 & 0 & \frac{1}{\sqrt{2}} & \frac{1}{\sqrt{2}} & 0\\
    0 & 0 & 0 & 0 & 1
  \end{pmatrix}
\end{equation}
and
\begin{equation}
  {\cal M}=
  \begin{pmatrix}
    M_1 & 0 & -g_1(v_u+v_d)/2 & g_1(v_u-v_d)/2 & 0 \\
    & M_2 & g_2(v_u+v_d)/2 & -g_2(v_u-v_d)/2 & 0 \\
    & & \mu_{\text{eff}} & 0 & -\lambda(v_u-v_d)/\sqrt{2} \\
    & & & -\mu_{\text{eff}} & -\lambda(v_u+v_d)/\sqrt{2} \\
    & & & & 0
  \end{pmatrix}.
\end{equation}
We split the mass matrix into the diagonal part and the off-diagonal
part, ${\cal M}={\cal M}^D+{\cal M}^O$, where ${\cal
  M}^D={\text{diag}}(M_1,\,M_2,\,\mu_{\text{eff}},\,-\mu_{\text{eff}},0)$
and ${\cal M}^O$ is the rest.
In the leading order, we have
\begin{equation}
  V^{(1)}_{nm}=\frac{{\cal M}^O_{mn}}{{\cal M}^D_{nn}-{\cal M}^D_{mm}}.
\end{equation}
In the perturbative limit, we find that
{\allowdisplaybreaks\begin{align}
  V^{(1)}_{51}&=-\frac{{\cal M}^O_{15}}{M_1}=0,\\
  V^{(1)}_{52}&=-\frac{{\cal M}^O_{25}}{M_2}=0,\\
  V^{(1)}_{53}&=-\frac{{\cal M}^O_{35}}{\mu_{\text{eff}}}=\frac{\lambda v(\sin\beta-\cos\beta)}{\sqrt{2}\mu_{\text{eff}}},\\
  V^{(1)}_{54}&=\frac{{\cal M}^O_{45}}{\mu_{\text{eff}}}=-\frac{\lambda v(\sin\beta+\cos\beta)}{\sqrt{2}\mu_{\text{eff}}}.
\end{align}
In the second order,}
\begin{equation}
  V^{(2)}_{nm}=\sum_{k\neq n} \frac{{\cal M}^O_{mk}{\cal
      M}^O_{nk}}{({\cal M}^D_{nn}-{\cal M}^D_{mm})({\cal
      M}^D_{nn}-{\cal M}^D_{kk})}-\frac12\sum_{k\ne n}\frac{|{\cal
      M}^O_{nk}|^2}{({\cal M}^D_{nn}-{\cal M}^D_{kk})^2}\delta_{mn}.
\end{equation}
Then, we have
{\allowdisplaybreaks \begin{align}
  V^{(2)}_{51}&=\frac{1}{M_1}\sum_{k\neq5}\frac{{\cal M}^O_{1k}{\cal M}^O_{5k}}{{\cal M}^D_{kk}}\nonumber\\
  &=
  \frac{1}{M_1}\biggl[\frac{g_1\lambda(v_u^2-v_d^2)}{2\sqrt{2}\mu_{\text{eff}}}+\frac{g_1\lambda(v_u^2-v_d^2)}{2\sqrt{2}\mu_{\text{eff}}}\biggr]
  =-\frac{g_1\lambda v^2 \cos2\beta}{\sqrt{2}M_1\mu_{\text{eff}}},\\
  V^{(2)}_{52}&=\frac{1}{M_2}\sum_{k\neq5}\frac{{\cal M}^O_{2k}{\cal M}^O_{5k}}{{\cal M}^D_{kk}}\nonumber\\
  &=\frac{1}{M_2}\biggl[-\frac{g_2\lambda(v_u^2-v_d^2)}{2\sqrt{2}\mu_{\text{eff}}}-\frac{g_2\lambda(v_u^2-v_d^2)}{2\sqrt{2}\mu_{\text{eff}}}\biggr]
  =\frac{g_2\lambda v^2 \cos2\beta}{\sqrt{2}M_2\mu_{\text{eff}}},\\
  V^{(2)}_{53}&=\frac{1}{\mu_{\text{eff}}}\sum_{k\neq5}\frac{{\cal M}^O_{3k}{\cal M}^O_{5k}}{{\cal M}^D_{kk}}=0,\\
  V^{(2)}_{54}&=-\frac{1}{\mu_{\text{eff}}}\sum_{k\neq5}\frac{{\cal M}^O_{4k}{\cal M}^O_{5k}}{{\cal M}^D_{kk}}=0.
\end{align}
Keeping all these terms, we can write}
\begin{eqnarray}
  &&N^{(1)}_{51}+N^{(2)}_{51}=V^{(1)}_{51}+V^{(2)}_{51}=-\frac{g_1\lambda v^2 \cos2\beta}{\sqrt{2}M_1\mu_{\text{eff}}},
  \label{eq:N_11_(1)(2)}\\
  &&N^{(1)}_{52}+N^{(2)}_{52}=V^{(1)}_{52}+V^{(2)}_{52}=\frac{g_2\lambda v^2 \cos2\beta}{\sqrt{2}M_2\mu_{\text{eff}}},
  \label{eq:N_12_(1)(2)}\\
  &&N^{(1)}_{53}+N^{(2)}_{53}=\frac{1}{\sqrt{2}}(V^{(1)}_{53}+V^{(1)}_{54})+\frac{1}{\sqrt{2}}(V^{(2)}_{53}+V^{(2)}_{54})=-\frac{\lambda v\cos\beta}{\mu_{\text{eff}}},
  \label{eq:N_13_(1)(2)}\\
  &&N^{(1)}_{54}+N^{(2)}_{54}=\frac{1}{\sqrt{2}}(-V^{(1)}_{53}+V^{(1)}_{54})+\frac{1}{\sqrt{2}}(-V^{(2)}_{53}+V^{(2)}_{54})=-\frac{\lambda v\sin\beta}{\mu_{\text{eff}}}.
  \label{eq:N_14_(1)(2)}
\end{eqnarray}
For large $\tan\beta$, or more precisely, for $\lambda
v\tan\beta/\mu_{\text{eff}}>1$, $N_{13}$ is very suppressed up to the
second order.
Thus, we include the third or higher order. The third order relation
is given as
% \begin{equation}
%   \begin{split}
{\allowdisplaybreaks\begin{align}
    V^{(3)}_{nm}={}&-\sum_{p,q\neq n}\frac{{\cal M}^O_{mp}{\cal M}^O_{pq}{\cal M}^O_{qn}}{({\cal M}^D_{mm}-{\cal M}^D_{nn})({\cal M}^D_{nn}-{\cal M}^D_{pp})({\cal M}^D_{nn}-{\cal M}^D_{qq})}\\
    &-\sum_{p\neq n}\frac32\frac{{\cal M}^O_{mn}|{\cal M}^O_{np}|^2}{({\cal M}^D_{nn}-{\cal M}^D_{mm})^2({\cal M}^D_{nn}-{\cal M}^D_{pp})}\\
    &-\sum_{p,q\neq n}\frac{{\cal M}^O_{np}{\cal M}^O_{pq}{\cal
        M}^O_{qn}}{({\cal M}^D_{nn}-{\cal M}^D_{pp})^2({\cal
        M}^D_{nn}-{\cal M}^D_{qq})}\delta_{nm}.
\end{align}
%   \end{split}
% \end{equation}
We here show only $V^{(3)}_{53}$ and $V^{(3)}_{54}$ since the other
third order terms are negligible.}
In the third order, we have
\begin{eqnarray}
  V^{(3)}_{53}&=&-\frac{\lambda v^3}{2\sqrt{2}\mu_{\text{eff}}^2}\biggl(\frac{g_1^2}{M_1}+\frac{g_2^2}{M_2}\biggr)\cos2\beta(\sin\beta+\cos\beta)\nonumber\\
  &&+\frac{3\lambda^3v^3}{2\sqrt{2}\mu_{\text{eff}}^3}\sin2\beta(\sin\beta-\cos\beta),
  \label{eq:N_13_(3)}\\
  V^{(3)}_{54}&=&-\frac{\lambda v^3}{2\sqrt{2}\mu_{\text{eff}}^2}\biggl(\frac{g_1^2}{M_1}+\frac{g_2^2}{M_2}\biggr)\cos2\beta(\sin\beta-\cos\beta)\nonumber\\
  &&+\frac{3\lambda^3v^3}{2\sqrt{2}\mu_{\text{eff}}^3}\sin2\beta(\sin\beta+\cos\beta),
  \label{eq:N_14_(3)}
\end{eqnarray}
and
\begin{eqnarray}
  &&N^{(3)}_{53}=-\frac{\lambda v^3}{2\mu_{\text{eff}}^2}\biggl(\frac{g_1^2}{M_1}+\frac{g_2^2}{M_2}\biggr)\cos2\beta\sin\beta+\frac{3\lambda^3v^3}{\sqrt{2}\mu_{\text{eff}}^3}\sin2\beta\sin\beta,\\
  &&N^{(3)}_{54}=-\frac{\lambda v^3}{2\mu_{\text{eff}}^2}\biggl(\frac{g_1^2}{M_1}+\frac{g_2^2}{M_2}\biggr)\cos2\beta\cos\beta+\frac{3\lambda^3v^3}{\sqrt{2}\mu_{\text{eff}}^3}\sin2\beta\cos\beta.
\end{eqnarray}
Up to now, we have not sorted the eigenstates by magnitudes of eigenvalues.
In the PQ-NMSSM, the singlino-like state is the lightest one in most parameter space.
Therefore, we should change the mixing matrix in order to fit with the
SUSY Les Houches Accord 2 (SLHA2) convention~\cite{Allanach:2008qq},
i.e. $N_{5i}\rightarrow N_{1i}$.

\section{Higgs-bottom quark and Higgs-gauge boson couplings}
The CP-even Higgs mass matrix (\ref{CPeven}) can be diagonalized
by the mixing matrix $S$ according to the SLHA2 convention.
\begin{equation}
  H_i^{\text{mass}}=S_{ij}H_j^{\text{weak}},
\end{equation}
where $H_i^{\text{weak}}=(H_{dR},\,H_{uR},\,S_R)$ and
$H_i^{\text{mass}}$ are ordered in increasing mass. Then the various
Higgs couplings can be expressed by the components of the
mixing matrix.
\begin{eqnarray}
  H_ib_Lb_R^c:&&
  \frac{y_b}{\sqrt{2}}S_{i1},\label{coupling:Hbb}\\
  H_iZ_{\mu}Z_{\nu}:&&
  g_{\mu\nu}\frac{v(g_1^2+g_2^2)}{\sqrt{2}}(S_{i1}\cos\beta+S_{i2}\sin\beta),
  \label{coupling:HZZ}\\
  H_iW_{\mu}^+W_{\nu}^-:&&
  g_{\mu\nu}\frac{vg_2^2}{\sqrt{2}}(S_{i1}\cos\beta+S_{i2}\sin\beta).
  \label{coupling:HWW}
\end{eqnarray}

%\section{Higgs-gluon and Higgs-photon couplings}

\section{Higgs-neutralino couplings}
\label{sec:Higgs-neut coup.}
Referring to Appendix A in \cite{Ellwanger:2009dp}, the
Higgs-neutralino couplings in the PQ-NMSSM are given by
\begin{equation}
  H_a\widetilde{\chi}_i^0\widetilde{\chi}_j^0~:~\frac{\lambda}{\sqrt{2}}(S_{a1}\Pi_{ij}^{45}+S_{a2}\Pi_{ij}^{35}+S_{a3}\Pi_{ij}^{34})
  +\frac{g_1}{2}(S_{a1}\Pi_{ij}^{13}-S_{a2}\Pi_{ij}^{14})-\frac{g_2}{2}(S_{a1}\Pi_{ij}^{23}-S_{a2}\Pi_{ij}^{24}),
\end{equation}
where
\begin{equation}
  \Pi_{ij}^{ab}=N_{ia}N_{jb}+N_{ib}N_{ja}. \nonumber
\end{equation}
Each term can be explained in piecewise.
The first term denotes $a$th Higgs coupling to $i$th and $j$th
neutralinos through the $H_{dR}$ component of the $a$th Higgs with the
$\widetilde{H}^0_u$ and $\widetilde{S}$ components of the neutralinos,
i.e. $\lambda H_{dR} \widetilde{H}^0_u \widetilde{S}/\sqrt{2}$.
The second and third terms are from permutations of the first term, i.e. $\lambda {H}_{uR} \widetilde{H}^0_{d} \widetilde{S} /\sqrt{2} $ and $\lambda S_{R} \widetilde{H}^0_{d} \widetilde{H}^0_{u}/\sqrt{2} $.
The fourth and fifth terms are from the SUSY U(1)$_Y$ gauge couplings
of $H_d$ and $H_u$ multiplets.
Likewise, the last two terms are from the SUSY SU(2)$_L$ gauge
interactions.

\end{document}